\documentclass[twocolumn,floatfix,prl,aps,showpacs]{revtex4-1}
\usepackage{graphicx,color,amsmath,amssymb}
\def\bc{\begin{center}}
\def\ec{\end{center}}
\def\beq{\begin{equation}}
\def\eeq{\end{equation}}
\def\bw{\begin{widetext}}
\def\ew{\end{widetext}}
\def\bea{\begin{eqnarray}}
\def\eea{\end{eqnarray}}

\renewcommand{\vec}[1]{{\textbf{\textit{#1}}}}

\usepackage{dcolumn}
\usepackage{bm}
\usepackage[caption=false]{subfig}
\usepackage{verbatim}

\begin{document}

\title{Competing Crystal Phases in the Lowest Landau Level}
\author{Alexander C. Archer,$^1$ Kwon Park,$^2$ and Jainendra K. Jain$^1$}
\affiliation{$^1$Department of Physics, 104 Davey Lab, The Pennsylvania State University, University Park, Pennsylvania 16802}
\affiliation{$^2$School of Physics, Korea Institute for Advanced Study, Seoul 130-722, Korea}

\date{\today}

\begin{abstract}
{We show that the solid phase between the 1/5 and 2/9 fractional quantum Hall states arises from an extremely delicate interplay between type-1 and type-2 composite fermion crystals, clearly demonstrating its nontrivial, strongly correlated character. We also compute the phase diagram of various crystals occurring over a wide range of filling factors, and demonstrate that the elastic constants exhibit non-monotonic behavior as a function of the filling factor, possibly leading to distinctive experimental signatures that can help mark the phase boundaries separating different kinds of crystals.}
\end{abstract}

\maketitle

A Wigner crystal~\cite{Wigner} (WC) is expected to form when the interaction energy of electrons dominates their kinetic energy. One way to accomplish this is to force all electrons in two dimensions into the lowest Landau level (LL) by applying a large magnetic field~\cite{Soviet}. The insulating phase at filling factors $\nu<1/6$ has been interpreted in terms of such a crystal \cite{Jiang90,Goldman90,Pan03,engel,ccli,ye,Chen2,Samban,Chen3}, although a definitive observation of the crystalline order is so far lacking. Remarkably, an insulating phase also appears between the fractional-quantum-Hall-effect~\cite{Tsui} (FQHE) liquids at $\nu=2/9$ and 1/5~\cite{Jiang90,Goldman90,Pan03}. The facts that this insulator has persisted even as the sample mobility has risen ten-fold, and that it is flanked by two FQHE liquids, suggest that the insulating behavior is probably caused by pinning of a crystal rather than individual carrier freeze-out. While a qualitative scenario for the re-entrant behavior can be constructed in terms of cusps in the energy of the liquid state \cite{Jiang90}, this behavior so far has not been explained by a quantitative theoretical calculation. We show in this paper that this insulating state results from an extremely subtle competition between the crystal and liquid states. Our results support the interpretation of this insulator as a pinned crystal, while also demonstrating its non-trivial nature as a crystal of composite fermions (CFs). We also consider the phase diagram of the crystal phase in a wider range of filling factors, calculate the elastic constants and predict their non-monotonic behavior as a function of $\nu$.

Numerous theoretical studies have considered the crystal phase~\cite{MZ,Lam,Lev,Esfar,Cote,Yi,Nare,Japan,Kun,man,Chang,he,Chang2,Jeon,QD2}. Maki and Zotos~\cite{MZ} (MZ) considered an uncorrelated Hartree-Fock WC of electrons in the lowest LL and evaluated its elastic properties. Lam and Girvin~\cite{Lam} (LG) considered a correlated WC, the energy of which has been compared \cite{Lam,JKam} with those of $1/m$ \cite{lau} and $n/(2pn+1)$ FQHE states \cite{Jain} ($m$ odd integer; $n$, $p$ integers), which shows a level crossing transition at $\nu\approx 1/6$. Beginning with Yi and Fertig~\cite{Yi}, a number of studies considered crystals of composite fermions~\cite{Nare,man,Chang,Chang2,Jeon,QD2}. In particular, Chang {\em et al.}~\cite{Chang} demonstrated that the CF crystals (CFCs) accurately capture the correlations in the crystal phase. 

For the questions addressed in this work, we need the energies of both the crystal and the FQHE states as a {\em continuous} function of $\nu$. For this purpose, we will consider two types of CFCs. Denoting composite fermions carrying $2p$ vortices by $^{2p}$CFs, these are:  (i) ``Type-1 $^{2p}$CFC" refers to a state in which {\em all} $^{2p}$CFs form a crystal. When pinned by disorder, this state will exhibit insulating behavior with divergent longitudinal resistance.  (ii) The term ``type-2 $^{2p}$CFC" refers to a state in which the excess CF particles or holes \cite{CFtext} relative to a FQHE liquid form a crystal. A type-2 CFC rides on the background of a FQHE liquid. In the presence of some disorder that pins the type-2 CFC, this state exhibits quantized Hall resistance and dissipationless transport. Type-2 CFCs, which can be likened to a pinned Abrikosov vortex lattice in a type-2 superconductors, are unobservable in transport experiments, but can be detected in microwave resonance experiments~\cite{Zhu} or by direct measurement of the spatial density profile (shown below for some cases).

We will consider $N$ electrons on the surface of a sphere exposed to a total flux $2Q$ in units of $hc/e$. This geometry~\cite{Haldane} is convenient for its lack of boundaries and obviates the complications requiring the introduction of ``ghost charges"~\cite{Yi}. We will denote the electron coordinates on the sphere as $\vec{r}_j=(\theta_j,\phi_j)$, $j=1,\cdots N$, and the crystal sites by $\vec{R}_l=(\gamma_l,\delta_l)$, with $l=1,\cdots N_c$, where $N_c$ is the number of lattice sites. It is also convenient to define the spinor variables $(u,v)=(\cos(\theta/2)e^{i\phi/2},\sin(\theta/2)e^{-i\phi/2})$ and $(U,V)=(\cos(\gamma/2)e^{i\delta/2},\sin(\gamma/2)e^{-i\delta/2})$. 
A problem with this geometry is that it is not possible to tile the surface of a sphere with a crystal without introducing defects. We place the crystal wave packet centers at the locations that minimize the energy of charged point particles on the surface of a sphere. Finding these locations, widely known as the Thomson problem~\cite{Thom}, has been accomplished previously by a number of researchers using powerful numerical techniques~\cite{Wales}. The Thomson crystal is locally a triangular WC, and the fraction of defects vanishes as $N_c\rightarrow \infty$.

We construct explicit wave functions as follows. The system of electrons at $2Q$ maps into a system of composite fermions at $2Q^*=2Q-2p(N-1)$ \cite{Jain}. We first construct $\Phi^\textrm{type-K-MZ}_{2Q^*, \{ \vec{R} \}}$, namely the type-K uncorrelated ``MZ crystal" at $2Q^*$ in which $N_c$ particles are located at the Thomson positions $\{\vec{R}_1,\cdots,\vec{R}_{N_c}\}$. We 
then obtain the type-K $^{2p}$CFC according to the mapping:
\begin{equation}
\Psi^\textrm{type-K-CFC}_{2Q, \{\vec{R}\}} =  {\cal P}_\textrm{LLL}   \prod_{j<k}(u_jv_k-v_ju_k)^{2p} \Phi^\textrm{type-K-MZ}_{2Q^*, \{ \vec{R} \}}
\label{CFC}
\end{equation}
where the Jastrow factor $ \prod_{j<k=1}^N(u_jv_k-v_ju_k)^{2p}$ attaches $2p$ vortices to electrons and 
${\cal P}_\textrm{LLL}$ is the lowest LL projection operator, which will be evaluated by the Jain-Kamilla method~\cite{JKam}.  Under this mapping, electrons, LLs and MZ crystals get converted into CFs, $\Lambda$Ls and CFCs, respectively.  For the type-1 $^{2p}$CFC, we have $N_c=N$ and the allowed $2p$ values are such that $N<2Q^*+1$; here 
\begin{align}
\Phi^\textrm{type-1-MZ}_{2Q^*, \{\vec{R}\}} =  \textrm{Det}\; \phi_{\vec{R}_l}^{2Q^*}\!\!(\vec{r}_j) =\textrm{Det}\;  (U_l^*u_j\!+\!V_l^*v_j)^{2Q^*}
\end{align}
where $\phi_{\vec{R}}^{2Q^*}\!\!(\vec{r})=(U^*u+V^*v)^{2Q^*}$ is the maximally localized wave packet in the lowest LL. 
For the type-2 $^{2p}$CFC, there are two possibilities. For certain values of $2Q^*$ we have $n\geq 1$ filled LLs and $N_c$ particles in the $(n+1)^{\rm th}$ partially filled LL. For convenience of illustration, let us take $n=1$, which corresponds to the FQHE state in the range $2/(4p+1)>\nu>1/(2p+1)$.  The wave function of the MZ crystal is given by:
\begin{align}
\Phi^\textrm{type-2-MZ}_{2Q^*, \{ \vec{R} \}} =\!
\left|
\begin{array}{ccc}
	Y_{Q^*Q^*-Q^*}(\!\vec{r}_1\!)& \ldots & Y_{Q^*Q^*-Q^*}(\!\vec{r}_N\!)\\
	\vdots & \ddots & \vdots\\
	Y_{Q^*Q^*Q^*}(\!\vec{r}_1\!) & \ldots & Y_{Q^*Q^*Q^*}(\!\vec{r}_N\!)\\
      {\cal O}^\dagger \phi_{\vec{R}_1}^{2(Q^*\!+1)}\!(\!\vec{r}_1\!) & \ldots & {\cal O}^\dagger \phi_{\vec{R}_1}^{2(Q^*\!+1)}\!(\!\vec{r}_N\!)\\
	\vdots & \ddots & \vdots\\
	{\cal O}^\dagger \phi_{\vec{R}_{N_c}}^{2(Q^*\!+1)}\!(\!\vec{r}_1\!) & \ldots & {\cal O}^\dagger \phi_{\vec{R}_{N_c}}^{2(Q^*\!+1)}\!(\!\vec{r}_N\!) 
\end{array} 
\right|
\end{align}
where $Y_{QQm}(\vec{r})\propto (-1)^{Q-m}v^{Q-m}u^{Q+m}$ are the lowest LL monopole harmonics~\cite{CFtext},
$N_c=N-(2Q^*+1)$, and
${\cal O}^\dagger=v^*\frac{\partial}{\partial u}-u^*\frac{\partial}{\partial v}$ is the LL raising operator in the spherical geometry~\cite{gre} (which also lowers the monopole strength by one unit). 
For $\nu<1/(2p+1)$, we form a crystal of $N_c=2Q^*+1-N$ CF-holes in the background of one filled $\Lambda$ level. Here the MZ crystal is obtained by taking the filled LL wave function $\prod_{j<k=1}^{2Q^*+1}(u_jv_k-v_ju_k)$ and replacing the $N_c$ coordinates $(u_j,v_j)$ with $j=N+1,\cdots, 2Q^*+1$ by 
the Thomson positions $(U_l,V_l)$. ${\cal P}_\textrm{LLL}$ is not required for either the type-2 CF-hole crystal or the type-1 CFC.

\begin{figure}[t]
\includegraphics[width=0.5\textwidth]{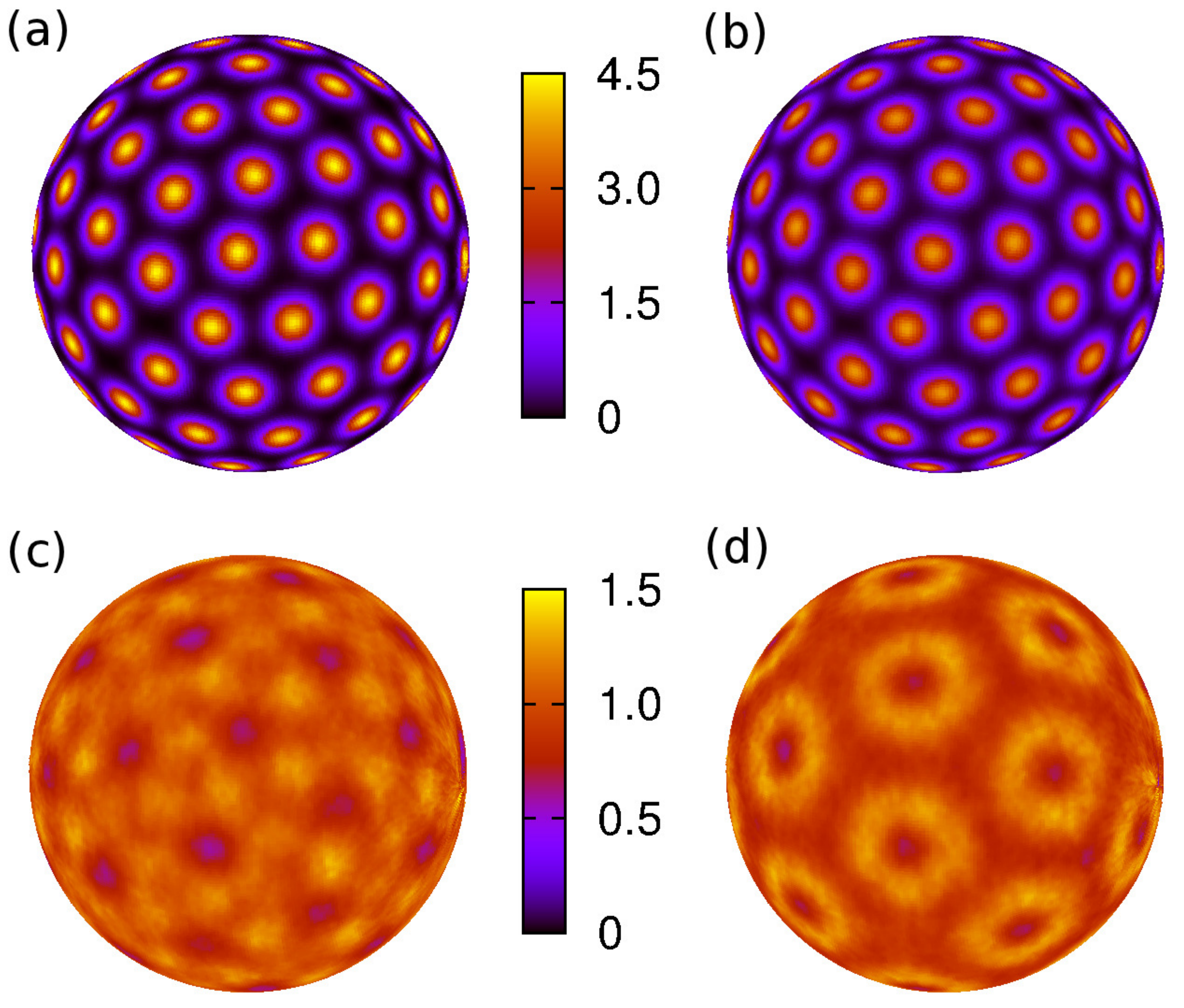}
\caption{
Density profiles for some type-1 and type-2 CF crystals on the surface of a sphere. All systems contain $N=96$ particles. The parameters are: (a) MZ crystal and (b) type-1 $^2$CFC for $2Q=433$ at $\nu=0.2188$; (c) type-2 $^4$CFC for $N_c=42$ and 
$2Q=433$ at $\nu=0.2188$; (d) type-2 $^4$CFC for $N_c=24$ and $2Q=451$ at $\nu=0.2103$. The filling factor in this region is determined using the interpolation relation $\nu= (N+2)/(2Q+15)$, which correctly reproduces the known finite size ``shifts" in $2Q$ at $\nu=1/5$ and 
$\nu=2/9$. The density is plotted in units of $\rho_0=N/4\pi R^2$. While comparing different plots, note that the radius of the sphere is $R=\sqrt{Q}$ in units of the magnetic length.
}
\label{fig:density_profile}
\end{figure}

Figure~\ref{fig:density_profile} shows the electron density profiles for various possible crystals for a filling factor slightly larger than 1/5: (a) type-1 MZ crystal, (b) type-1 $^2$CFC, and (c-d) type-2 $^4$CFC. Correlations in the type-1 $^2$CFC result in a  slight delocalization of the electrons at the lattice sites as compared to the MZ crystal (compare Figs. 1a and 1b). The type-2 CFC looks remarkably different. An isolated CF in the second $\Lambda$ level is known to have the shape of a ring~\cite{CFtext}. A ``ring crystal" is clearly seen in Fig.~\ref{fig:density_profile}(d) where the composite fermions in the second $\Lambda$L are far from one another. The lattice spacing decreases with increasing filling factor, and the rings begin to overlap producing a complex interference pattern as seen in Fig.~\ref{fig:density_profile}(c), in which the density maximum occurs on the line joining adjacent sites producing a ``bond crystal." Even more intricate density profiles can occur for type-2 CFCs in higher $\Lambda$Ls.

\begin{figure}[t]
\includegraphics[width=0.50\textwidth]{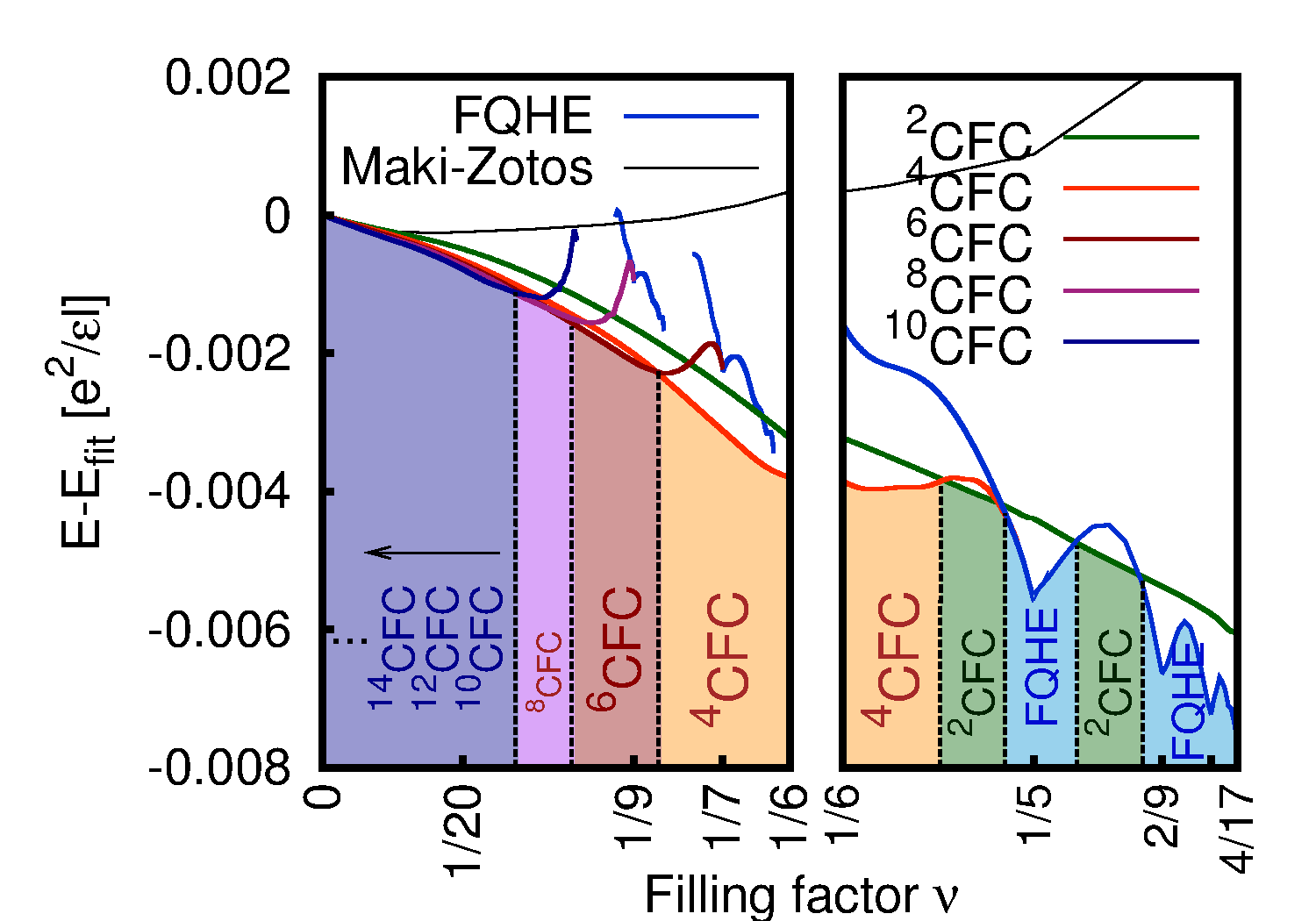}
\includegraphics[width=0.50\textwidth]{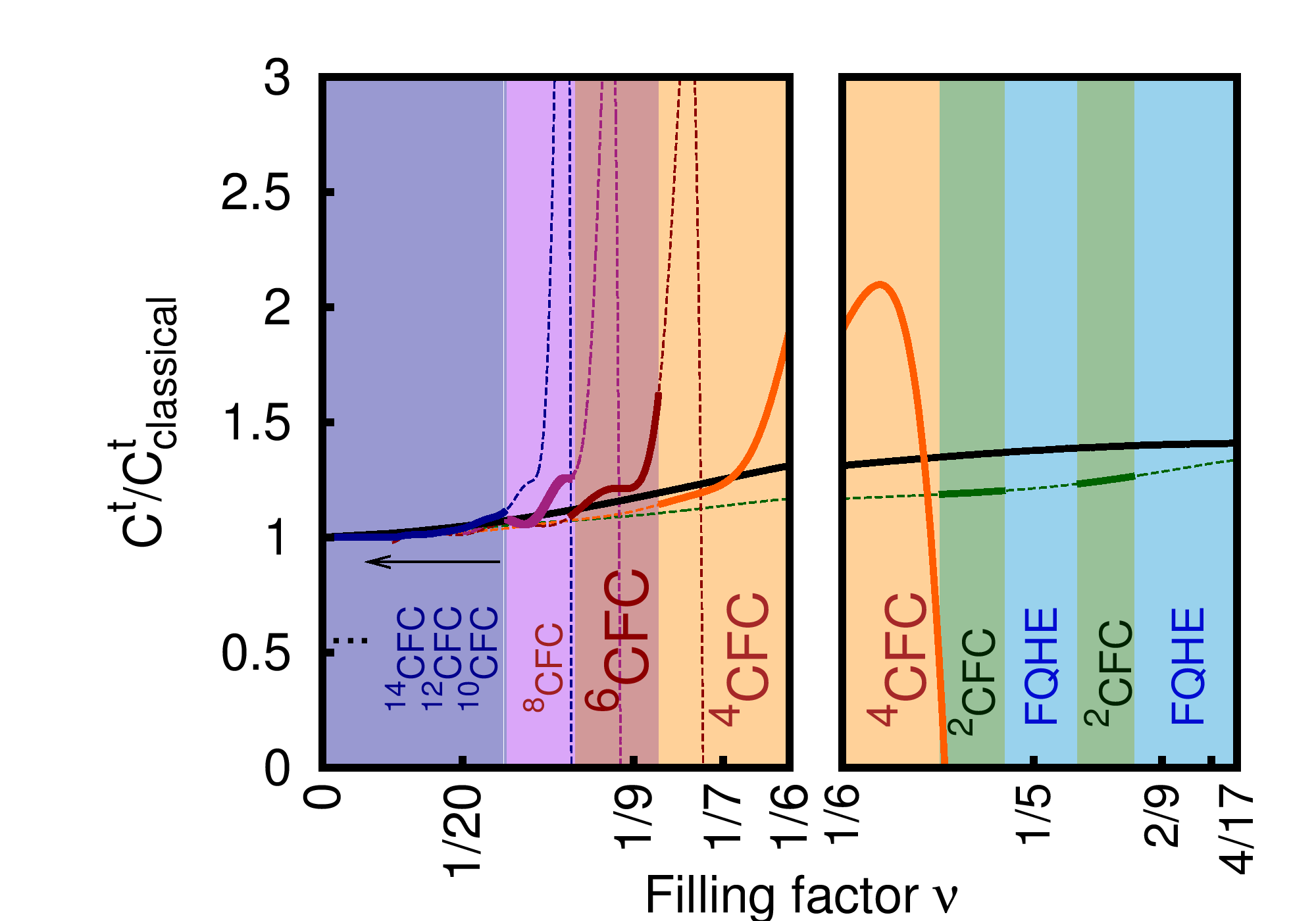}
\caption{
Upper panels: Energy per particle as a function of filling factor for various type-1 and type-2 CFC states. The latter are labeled FQHE. All energies are quoted relative to a reference energy $E_{\rm fit}=-0.782133\nu^{1/2}+0.2623\nu^{3/2}+0.18\nu^{5/2}-15.1e^{-2.07/\nu}$, which has a form similar to that in Ref.~\cite{Lam}  but with coefficients modified to display the energy differences between the competing states more clearly. Lower panels: Shear modulii of type-1 crystals of composite fermions with $2p$ vortices. The shear modulus of the MZ crystal, given by the solid black line, is shown for reference. The shear modulus of the $^{2p}$CFC is indicated by a solid line in the regime where it is the ground state, and by a dashed line otherwise.  The regions $0<\nu<1/6$ and $1/6<\nu<4/17$ are shown in separate panels because different filling factor scales are used for them.
}
\label{fig:Etot}
\end{figure}

Figure~\ref{fig:Etot} shows the energies, obtained via standard Metropolis Monte Carlo techniques~\cite{CFtext}, for several type-1 and type-2 CFCs (the latter labeled FQHE) for 96 particles as a function of $\nu$.  A re-entrant insulating phase appears in a filling factor range between the 1/5 and 2/9 FQHE states, where the type-1 CFC beats the FQHE state (supporting a type-2 CFC) by an energy of 0.0005 $e^2/\epsilon \ell$ per particle. To put this in perspective, we recall that the theoretical excitation gaps at 1/3 and 1/5 are $\sim$0.1 $e^2/\epsilon \ell$ and $\sim$0.025 $e^2/\epsilon \ell$, respectively. The MZ crystal does not produce the re-entrant crystal phase; in spite of apparently small differences in the density profiles, the energy of the MZ crystal is higher by $\sim$ 0.006 $e^2/\epsilon \ell$ per particle than the energy of the type-1 CFC at $\nu\approx 1/5$. The energy of the LG crystal is 0.00214 $e^2/\epsilon \ell$ above the 1/5 FQHE state and 0.00305 $e^2/\epsilon \ell$ above the 2/9 FQHE state (using the thermodynamic limits from Refs.~\cite{Lam,JKam}.), and will also not capture the insulating crystal phase between 1/5 and 2/9. The understanding of the insulating phase between 1/5 and 2/9 as the $^2$CF crystal leads to the intuitively pleasing picture in which the $^{4}$CFs of the nearby liquid states shed two of their vortices to establish a crystal, while retaining energetically favorable correlations through the remaining two vortices.

It has recently been demonstrated \cite{Muraki} that the density distribution of the electrons can be accessed through NMR measurements, because the Knight shift is proportional to the local electron density. As shown in the Supplemental Materials (SM) \cite{SM}, the type-1 and type-2 CFCs have remarkably different density distributions, which may allow NMR to identify the phase boundaries in the region $1/5 < \nu < 2/9$.

We have also studied the competition between the liquid and the crystal phases at lower fillings, where we consider type-1 $^{2p}$CFCs with different choices of $2p$ to determine which produces the lowest energy.  As $\nu$ is lowered below 1/5, a series of type-1 $^{2p}$CFCs with increasing vorticity occurs. No FQHE state supporting a type-2 CFC appears for $\nu<1/6$. (We cannot rule out FQHE states with fillings $\frac{n}{6n-1}$ in the range $1/5>\nu>1/6$, not studied here due to complications associated with reverse flux attachment~\cite{WuDevJain}.) The phase boundaries practically remain unchanged for $N>32$ (see SM), and thus represent the thermodynamic limit. 
We have also studied the effect of finite thickness, which does not change the phase diagram appreciably. 
Using the model of Ref.~[\onlinecite{Shi2}], we have considered quantum well structures with the well widths ranging from 20 to 80 nm and the electron density ranging from $1.0\times 10^{10}$ to $1.5\times10^{11}\,\text{cm}^{-2}$, and found that even though the energy per particle decreases by up to $10\%$ for the largest densities and widths considered, the phase boundaries are little changed.

To formulate the low-energy dynamics of type-1 CFCs, 
we begin by modeling the CFC as a collection of charged point particles that interact through an effective interaction $V(R_{jk})$ with $R_{jk}=| \vec{R}_j - \vec{R}_k |$, but are otherwise classical. 
The dynamical matrix $\Phi_{\alpha\beta}(\vec{k})$, where $\alpha,\beta$ denote spatial directions, is given by~\cite{MZ}:
\begin{align}
\displaystyle \Phi_{\alpha\beta}(\vec{k}) &= \sum_j \left[ 1-\cos(\vec{k} \cdot\vec{R}_j) \right]
\frac{\partial^2V(R_j)}{\partial R_{j, \alpha} \partial R_{j, \beta}}
\nonumber 
\\ 
&\simeq \left[ \frac{\nu}{k} + (C^{\rm L} -C^{\rm t} ) \right] k_{\alpha}k_{\beta} +\delta_{\alpha\beta} C^{\rm t} k^2
\label{eq:dmat}
\end{align}
where $R_j=|\vec{R}_j|$.
In the above, the second line is obtained in the $k\rightarrow0$ limit under the explicit assumption of the $C_6$ symmetry. Between the two elastic parameters, $C^{\rm L}$ and $C^{\rm t}$, the shear modulus $C^{\rm t}$ is of special importance because it determines the low-energy behavior of the magnetophonon mode and its becoming negative signals an instability of the crystal. As shown in the SM, for the hexagonal lattice it can be obtained directly from the energy per particle of the crystal by the following equation:
\begin{equation}\label{eq:cfsm}
C^{\rm t}_{\rm CF}= \displaystyle \frac{1}{2}\nu^2\frac{\partial^2}{\partial\nu^2}(E_{\rm CF}-E_{\rm MZ})+C^{\rm t}_{\rm MZ}
\end{equation}
where $E_{\rm MZ}$ is defined as the energy of a hexagonal crystal of {\em classical} particles interacting with the MZ interaction $V_{\rm MZ}=\frac{\sqrt{\pi}}{4}\frac{I_0(R^2/8)}{\cosh(R^2/8)}$, and the derivatives with respect to $\nu$ are to be evaluated at fixed $B$ (i.e. fixed $\ell$).
The derivation of this relation relies on the assumptions that the dynamics of the crystal can be described in the harmonic approximation and that the total energy can be approximated as a sum of two-body interactions. The latter approximation becomes unreliable as the system approaches $\nu^*=1$, or $\nu=1/(2p+1)$, where the crystal wave function actually merges into a FQHE liquid (this is analogous to the fact that at $\nu=1$ the MZ wave function describes the $\nu=1$ liquid state). However, for the MZ wave function, this assumption is quite accurate for $\nu<1/2$ (see SM), and by analogy, we expect it to be accurate for $^{2p}$CFC for up to $\nu^*<1/2$, or $\nu<1/(2p+2)$.

The lower panel of Fig.~\ref{fig:Etot} shows the shear modulus of the $^{2p}$CFCs as a function of filling factor. As noted above, the shear modulus for $^{2p}$CFC close to $\nu=1/(2p+1)$ is quantitatively unreliable due to the importance of the three and higher body terms in the effective interaction that have been neglected above. Fortunately, 
in the physically relevant regions, we have $\nu^*<1/2$, and therefore we believe that $C^{\rm t}_{\rm CF}$ shown by the solid lines is accurate; the only exception is for $^4$CFC for which the $C^{\rm t}_{\rm CF}$ for $\nu>1/6$ is not quantitatively reliable. The shear modulus exhibits, unlike for the MZ or LG crystal, a series of discontinuities at the phase boundaries, which serve as a possible way of measuring the phase diagram.

\begin{figure}
\includegraphics[width=0.45\textwidth]{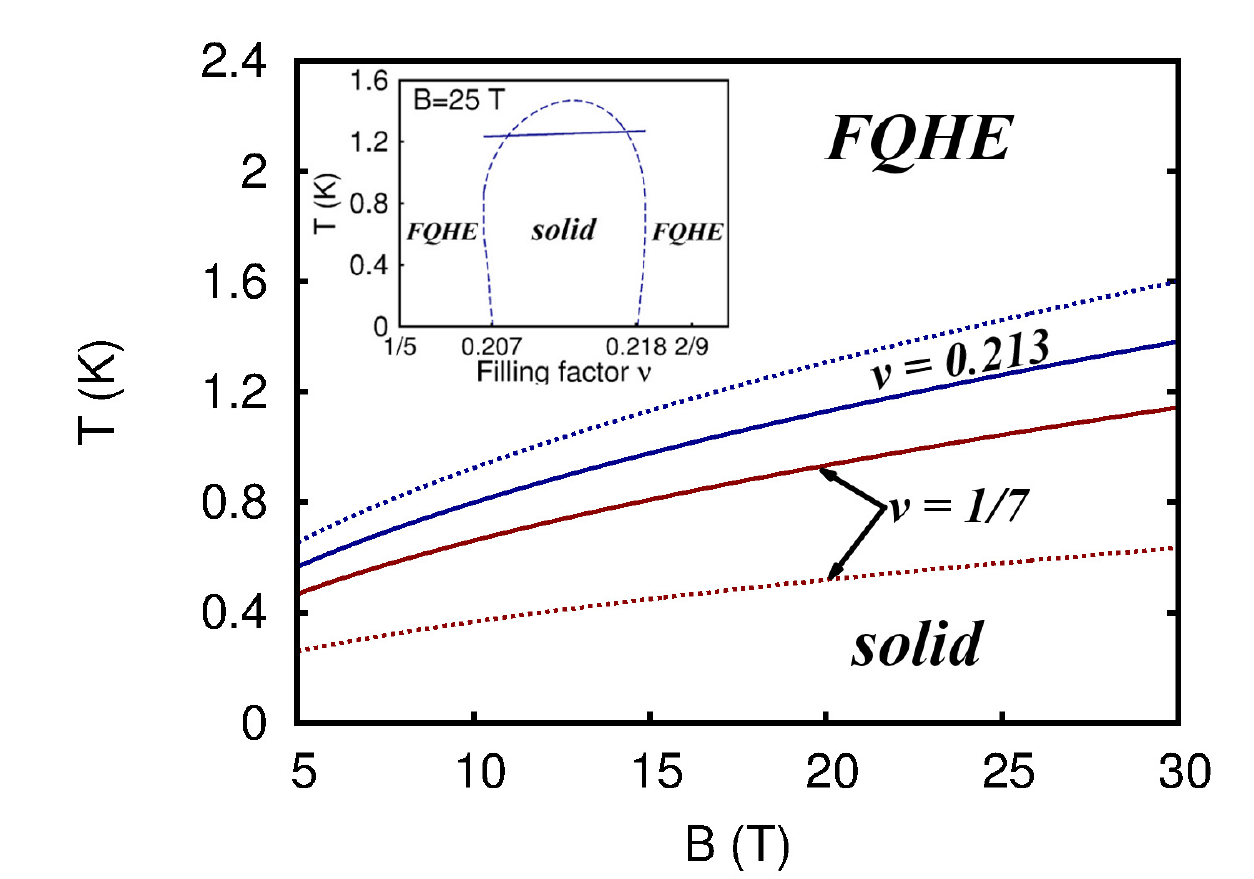}
\caption{Melting temperature of the CF crystal at $\nu=1/7$ and $\nu=0.213$ as determined by the Kosterlitz-Thouless mechanism (solid lines) and a first order transition into a FQHE liquid~\cite{Price} (dotted lines). The inset shows the phase diagram in the filling factor range $1/5\leq\nu\leq2/9$ for $B=25$ T.
}
\label{fig:phb}
\end{figure}

Transport~\cite{Jiang90,Goldman90,Pan03} and photoluminescence~\cite{buh,hayne,gold} experiments have probed the temperature dependence of the insulating phase. The melting of the crystal is of interest, and two possibilities can result in remarkably different experimental manifestations. The Kosterlitz-Thouless (KT)~\cite{Koster,Thouless} melting temperature is given by \cite{MZ}:
$k_{\rm B}T_{\rm KT} /(e^2/\epsilon \ell)= (2\pi\sqrt{3})^{-1}(C^{\rm t}_{\rm CF}/C^{\rm t}_{\rm classical}) 0.09775\nu^{1/2}$, where $C^{\rm t}_{\rm classical}=0.09775\nu^{1/2}$.
Another possibility is that of melting into the FQHE liquid, considered by Price {\em et al.}\cite{Price}, the transition temperature $T_{\rm M}$ for which is determined by the competition of the free energies of the crystal and liquid states \cite{SM}. The resulting transition temperatures are shown in Fig.~\ref{fig:phb}. We find that at $\nu=1/7$ (and also lower fillings) the transition occurs into a FQHE liquid, whereas for the crystal between $1/5<\nu<2/9$ the melting is governed by the KT physics; the difference arises because of much larger roton gap at 1/5. The inset shows the phase boundary as a function of the filling in the range $1/5<\nu<2/9$. Note that, for a narrow range of $\nu$, the FQHE state freezes into a type-1 CFC with increasing temperature and then melts back into the FQHE state; the possibility of a similar re-entrant transition was noted previously in Ref.~\cite{Price} at $\nu=1/3$ and 1/5 at certain values of LL mixing.

Chitra \emph{et al.} \cite{Chitra1,Chitra2} have developed an elastic model which predicts the pinning frequency of the classical WC. Using the shear moduli of the type-1 $^{2p}$CFCs, we find non-trivial quantum corrections to the classical pinning frequencies. The magnetic field dependence of the pinning frequency has two forms depending on the length scale of disorder, $r_{f}$:
$\omega_p\propto \nu/C^{\rm t}_{\rm CF}$ for $r_f>\ell$, and $\omega_p\propto (\nu^2 C^{\rm t}_{\rm CF})^{-1}$ for $r_f<\ell $, assuming constant density. Clearly, the discontinuity of the $C^{\rm t}_{\rm CF}$ at the phase boundaries will translate into a discontinuity in $\omega_p$.

Several features of our calculation are consistent with experimental observations. The range of the re-entrant crystal in Fig.~2, $0.207 < \nu < 0.218$, agrees with the region where activated behavior has been observed \cite{Jiang90}. No type-1 $^{2p}$CFC appears for $\nu>2/9$, consistent with an absence of an insulator here in high quality samples. The observation of FQHE-like structure at very low fillings (such as 1/7 and 1/9) observed in Ref.~\cite{Pan03} at somewhat elevated temperatures is consistent with a melting of the crystal into a FQHE state. The frequency dependent conductivity measured in microwave absorption experiments shows resonances between 1/5 and 2/9 \cite{Chen3}, which continue in the insulator below 1/5 until $\nu=0.18$. 
It is tempting to attribute these features to the $^2$CFC on either side of the 1/5 state. A qualitative change in the behavior and a decreasing pinning frequency below $\nu=0.18$ may indicate a transition into a $^4$CFC, although further work will be needed for a conclusive statement. We note that unlike for MZ or LG crystals, the pinning frequency of the CFC is predicted to have a complicated dependence on $\nu$ and can sometimes decrease with increasing $\nu$ in the regime $r_f>\ell$; such behaviors have been seen in lower mobility samples  \cite{engel,ccli}. The transition temperatures obtained above are generally higher than those estimated from experiments (although a clean transition has not yet been observed).  We also note that we have not considered disorder and LL mixing, which will affect the phase boundaries and melting temperatures.

A. C. A. and J. K. J. thank the National Science Foundation for support under Grant No. DMR-1005536 and the Research Computing and Cyberinfrastructure, a unit of Information and Technology Services at Pennsylvania State University for providing high-performance computing resources. K. P. thanks the National Research Foundation of Korea (NRF) funded by the Korea government (MSIP) under Quantum Metamaterials Research Center Grant No. 2008-0062238.

\clearpage

\begin{widetext}

\begin{appendix}
 \section{\large{Supplementary Material for ``Competing Crystal Phases in the Lowest Landau Level''}}

\section{S1: Free energies of the crystal and liquid states}

The free energy of the type-1 $^{2p}$CFC at small temperatures is given by [\onlinecite{pr1}]:
\begin{equation}\label{eq:fws1}
 F^{\rm CFC}=E^{\rm CFC}-0.701\left(\frac{C^{\rm t}_{\rm classical}}{C^{\rm t}_{\rm CF}}\frac{2a_0}{\nu e^2/\epsilon}\right)^{4/3}\left(k_{\rm B}T\right)^{7/3}
\end{equation}
where $E^{\rm CFC}$ is the $T=0$ energy per particle of the CFC and $a_0$ is the lattice spacing. The free energy of the FQHE liquid is given by  [\onlinecite{pr1}]:
\begin{equation}
 F^{\rm FQHE}=E^{\rm FQHE}-(2\pi m_{\rm R}\ell^2)^{1/2}\frac{k_{\rm R}\ell}{\nu}(k_{\rm B}T)^{3/2}e^{-\Delta_{\rm R}/k_{\rm B}T}
\end{equation}
where $E^{\rm FQHE}$ is the $T=0$ energy per particle of the FQHE state with a type-2 $^4$CFC, where it is assumed that the temperature is sufficiently small that only the magnetoroton [\onlinecite{Girvin}] part of the dispersion is relevant; $\Delta_{\rm R}$ and $m_{\rm R}$ are the energy and mass of the magnetoroton, and $k_{\rm R}$ is the wave vector at the energy minimum. In our calculation of the temperature where the $^2$CF crystal melts into a FQHE state at $\nu=0.213$, we have assumed that the type-2 CFC is pinned and the lowest energy excitations are the magnetorotons of the 1/5 state. 
The parameters $\Delta_{\rm R}$, $m_{\rm R}$ and $k_{\rm R}$ are obtained by performing fits to the CF exciton dispersion in the vicinity of the lowest energy magnetoroton as shown in Fig.~\ref{fig:fit}. For the crystal states at $\nu=1/5$ and 1/7, we have $C^{\rm t}_{\rm CF}/C^{\rm t}_{\rm classical}\approx1.2$.

\section{S2: System size dependence of energies}
In Fig.~\ref{fig:totE} we show the energies of type-1 $^{2p}$CFCs as a function of the system size. For clarity, we mark the phase boundaries between the crystals with a thin vertical line and the filling factor of the transition. This figure shows that the type-1 $^{2p}$CFC phase boundaries are well converged for system sizes greater than 32 particles.

Fig.~\ref{fig:ediff} shows the energy difference between the type-1 $^2$CFC and the FQHE state with a type-2 $^4$CFC in the filling factor range $1/5\leq\nu\leq2/9$ for three different system sizes. (The typical error in the energy difference from Metropolis sampling is $\sim 3\times10^{-5} e^2/\epsilon \ell$.) This shows that the results for 96 particles accurately represent the thermodynamic limit. 

\section{S3: Calculating the Shear Modulus of the $^{2p}$CF crystals}

In this section we first show how the shear modulus can be obtained from the filling factor dependence of the energy per particle of the hexagonal lattice, under the assumptions that the dynamics of the crystal can be described in the harmonic approximation and that the total energy can be written as a sum of two-body interactions. The derivation proceeds along two steps. In the first step, we show how to calculate the shear modulus from the dynamical matrix. In the second step, we show that the second derivative of the energy per particle with respect to the filling factor is proportional to the shear modulus.  We then provide details of our procedure for obtaining the shear modulus from our finite system results for which the energy is known only at discrete values of $\nu$.

In what follows, we will set the magnetic length $\ell=1$. The variation in filling factor $\nu$ should be thought of in terms of a change in density while keeping the magnetic field $B$ constant. The hexagonal lattice sites are located at
\begin{equation}
\vec{R}_{n,m}=\sqrt{\frac{4\pi}{\sqrt{3}\nu}}\left(n+{m\over 2},\frac{\sqrt{3}}{2}m\right).
\end{equation}

The dynamical matrix, $\Phi_{\alpha\beta}(\vec{k})$, of the classical 2D hexagonal system exposed to a perpendicular magnetic field is given by\cite{bm,mz}:
\begin{equation}\label{eq:dynmat}
 \displaystyle\Phi_{\alpha\beta}(\vec{k})=\sum_{\vec{R}\neq0}\left(1-\cos(\vec{k}\cdot\vec{R})\right)\frac{\partial^2V(R)}{\partial R_{\alpha} \partial R_{\beta}}=\left(\frac{\nu}{k}+C^{\rm L}-C^{\rm t}\right)k_{\alpha}k_{\beta} + C^{\rm t}k^2\delta_{\alpha\beta} + O(k^3)
\end{equation}
where the far right hand side of Eq.~(\ref{eq:dynmat}) gives the small $k$ limit, $R=|\vec{R}|$, $C^{\rm L}$ and $C^{\rm t}$ are the elastic constants which characterize the hexagonal lattice and $C^{\rm t}$ is the shear modulus. One can extract the shear modulus directly from the lattice sum in Eq.~(\ref{eq:dynmat}):
\begin{equation}
\displaystyle \Phi_{xx}((0,k_y))=\sum_{\vec{R}\neq0}\left(1-\cos(k_yR_y)\right)\frac{\partial^2V(R)}{\partial R_x \partial R_x}=k_y^2C^{\rm t}
\end{equation}
The long range interaction between $^{2p}$CFCs is Coulombic, while their short-range behavior gives rise to interesting elastic properties. In order to capture their short range behavior and to eliminate convergence issues, we define $V'\equiv V-V_{\rm MZ}$:
\begin{equation}
 \displaystyle C^{\rm t}_{\rm CF}=\frac{1}{k_y^2}\sum_{\vec{R}\neq0}^{\vec{R}_{\rm upp}}\left(1-\cos(k_yR_y)\right)\frac{\partial^2V'(R)}{\partial R_x \partial R_x} + C^{\rm t}_{\rm MZ}
\end{equation}
where $\vec{R}_{\rm upp}$ is chosen to evaluate the sum to arbitrary accuracy, and $V_{\rm MZ}$ is the MZ interaction given by\cite{mz} 
\begin{equation}
V_{\rm MZ}(R)=\frac{\sqrt{\pi}}{4}\frac{I_0(R^2/8)}{\cosh(R^2/8)}
\end{equation}
which is the Coulomb energy of two properly antisymmetrized gaussian wave packets in the lowest Landau level at distance $R$. 
Next, we expand $\cos({k_yR_y})$ in a Taylor series around $k_y=0$ and then take the limit $k_y\rightarrow0$:
\begin{equation}\label{eq:sm3}
 C_{\rm CF}^{\rm t}=\frac{1}{2}\displaystyle\sum_{\vec{R}\neq0}^{\vec{R}_{\rm upp}}(R_y)^2\frac{\partial^2V'(R)}{\partial R_x \partial R_x} + C^{\rm t}_{\rm MZ}
\end{equation}
Using the following identities (which rely on the hexagonal symmetry of the lattice):
\begin{equation}
 \frac{\partial^2}{\partial R_x^2}=\frac{(R_y)^2}{R^3}\frac{\partial}{\partial R} + \frac{(R_x)^2}{R^2}\frac{\partial^2}{\partial R^2}
\end{equation}
\begin{equation}
 \displaystyle\sum_{\vec{R}\neq0}(R_y)^4f(R)=\frac{9}{24}\sum_{\vec{R}\neq0}R^4f(R)
\end{equation}
\begin{equation}
 \displaystyle\sum_{\vec{R}\neq0}(R_y)^2(R_x)^2f(R)=\frac{3}{24}\sum_{\vec{R}\neq0}R^4f(R)
\end{equation}
we can rewrite Eq.~(\ref{eq:sm3}) as:
\begin{equation}\label{eq:sm4}
 C^{\rm t}_{\rm CF} = \frac{1}{16}\displaystyle\sum_{\vec{R}\neq0}\left(3R\frac{\partial V'}{\partial R}+R^2\frac{\partial^2 V'}{\partial R^2}\right) + C^{\rm t}_{\rm MZ}
\end{equation}
This completes the first step.

We now show explicitly how to calculate the shear modulus from the energy per particle. We note the following identities (when applied on a function of $R$):
\begin{equation}
 \frac{\partial}{\partial \nu}=\frac{\partial R}{\partial \nu}\frac{\partial}{\partial R}=-\frac{1}{2\nu}R\frac{\partial}{\partial R}
\end{equation}
\begin{equation}
 \frac{\partial^2}{\partial \nu^2}
 =\frac{1}{\nu^2}\left(\frac{3}{4}R\frac{\partial}{\partial R} + \frac{1}{4}R^2\frac{\partial^2}{\partial R^2}\right)\label{eq:finr}
\end{equation}
where we have used that $R\sim \nu^{-1/2}$.
Applying Eq.~(\ref{eq:finr}) to the energy per particle, we arrive at the final result:
\begin{equation}\label{eq:cfsm}
 \displaystyle \frac{1}{2}\nu^2\frac{\partial^2}{\partial\nu^2}(E_{\rm CF}-E_{\rm MZ})+C^{\rm t}_{\rm MZ}=\frac{1}{4}\nu^2\frac{\partial^2}{\partial\nu^2}\sum_{\vec{R}\neq0}V(R)'+C^{\rm t}_{\rm MZ}=C^{\rm t}_{\rm CF}
\end{equation}
Here, $E_{\rm MZ}$ is defined as the energy of a hexagonal crystal of {\em classical} particles interacting with the MZ interaction $V_{\rm MZ}$.  

This relation can be explicitly verified for the shear modulus of a hexagonal lattice of classical particles. The classical energy per particle for the 2d planar hexagonal lattice\cite{bm,mz} is given by $E=-0.782133\nu^{1/2}$, the shear modulus is given by $C^{\rm t}_{\rm classical}=0.0978\nu^{1/2}$ and the two are related by:
\begin{equation}
 \frac{1}{2}\nu^2\frac{\partial^2}{\partial\nu^2}\left(-0.782133\nu^{1/2}\right)=0.0978\nu^{1/2}
\end{equation}
We have also tested that the energies obtained by MZ\cite{mz} in which they assume two-body interaction, and confirmed that Eq.~(\ref{eq:cfsm}) exactly reproduces the shear modulus.

We next come to the issue of extracting the shear modulus from the discrete data points available to us from our finite system studies. For this purpose, we fit a smooth analytical curve to the energy per particle and calculate the shear modulus using Eq.~(\ref{eq:cfsm}). 
We show our smooth analytical fits to the $^{2p}$CFC energy relative to the Maki-Zotos energy $E_{\rm MZ}$ and their corresponding shear moduli for $N=128$ particles in Fig.~\ref{fig:ct}. Here, the energy $E_{\rm MZ}$ is given by the total energy of 128 classical particles located at the Thomson minima on the sphere that interact through the Maki-Zotos two-body interaction $V_{\rm MZ}$. The black curve in the right column panels of Fig.~\ref{fig:ct} is the shear modulus calculated from the strictly two-body Maki-Zotos interaction and is included for reference. 

To fit the CF energy, we use the functional form for the curves:
\begin{equation}
 \displaystyle f^{2p}_{\rm i}(\nu)=\left(\sum_{j=0}^{\rm m_{2p}+i}a_{2p,i}^j\nu^{3/2+j}\right)e^{-\nu^2}
\end{equation}
where $a_{2p,i}^j$ are found using a Least Squares fitting routine in Mathematica and $i$ is the fit number: $i=1,2,3$ or 4. The lines in Fig.~\ref{fig:ct} are drawn using these fit functions. In each fit function, referred to as `Fit $i$' in Fig.~\ref{fig:ct}, the degree of the polynomial is given by $m_{2p}+i$. The value of $m_{2p}$ is chosen separately for each crystal to give the best fit result. We perform fits of varying degree in order to check the consistency of our results, and take as our final result the average of four fits (given by the solid purple line in the right column panels of Fig.~\ref{fig:ct}). The weak oscillations in the shear modulus calculations are a result of the high degree of the polynomial fit. We note that the oscillations in the shear modulus fits are much smaller than the large overall structure that we observe near the transitions between $^{2p}$CFCs, implying that the latter is not an artifact of the fitting procedure. We have performed fits over the entire filling factor range that each $^{2p}$CFC exists, but have only shown the fits and their corresponding shear modulus for $\nu^*<1/2$ (or $\nu<{1\over 2p+2}$).

As a benchmark, in Fig.~\ref{fig:ctp0} we compare our calculated shear modulus for the $^0$CFC, which is a Slater determinant of otherwise uncorrelated electron wave packets, with that obtained previously by MZ. The two curves are very similar, but not identical. The difference arises because while the $^0$CFC is a fully antisymmetrized wave function, the MZ calculation imposes only pairwise antisymmetry, which is a good approximation only for small $\nu$. We believe that the shear modulus obtained in our calculation is more reliable. Interestingly, while Maki-Zotos results shows an instability of the crystal for $\nu\approx 0.45$, where the shear modulus becomes negative, our calculation eliminates that instability.  Of course, both are subject to the approximation that neglects three and higher body terms in the effective model mapping it into a crystal of classical particles.
For other values of $2p$, both the modified interaction between composite fermions and the antisymmetrization contribute to the deviation from the MZ shear modulus. Given that the distance between the neighboring lattice sites is large compared to the magnetic length, we expect that the assumption that the interaction is dominated by 2-body terms is accurate for $\nu^*<1/2$ or $\nu<1/(2p+2)$. However, as $\nu^*$ begins to approach unity, this approximation clearly breaks down (at $\nu^*=1$ we do not even have a crystal, but a liquid). Fortunately, the physically relevant regions (i.e. where the shear modulus is shown by solid lines) correspond to  $\nu^*<1/2$, with the exception of the filling factor $1/6<\nu<1/5$.

\section{S4: Density distribution of CF crystals}

Recently, Muraki and collaborators \cite{Muraki} have found that the spectral line shape of the NMR spectra contains information about the electron density distribution, because the Knight shift is proportional to the local electron density. We plot in Fig.~\ref{fig:pdens} the normalized probability of the CFC density as a function of density 
evaluated for a system with 96 particles. 
These density distributions are obtained at $\nu=0.215$ for type-1 crystals with $2p=0$ and $2$, and the type-2 crystal in the background of the $1/5$ FQHE state. From these plots it is clear that while the density distributions of the two type-1 CFCs are quite similar, they both qualitatively differ from the type-2 CFC, and it may be possible to identify the phase boundaries separating the type-1 and type-2 CFCs in the region $1/5<\nu<2/9$ by NMR measurements using the analysis of Ref. [\onlinecite{Muraki}]; in particular, an abrupt change in the dominant Knight shift is predicted at the phase transition.

\begin{figure}[h]
\begin{center}
\includegraphics[width=0.40\textwidth]{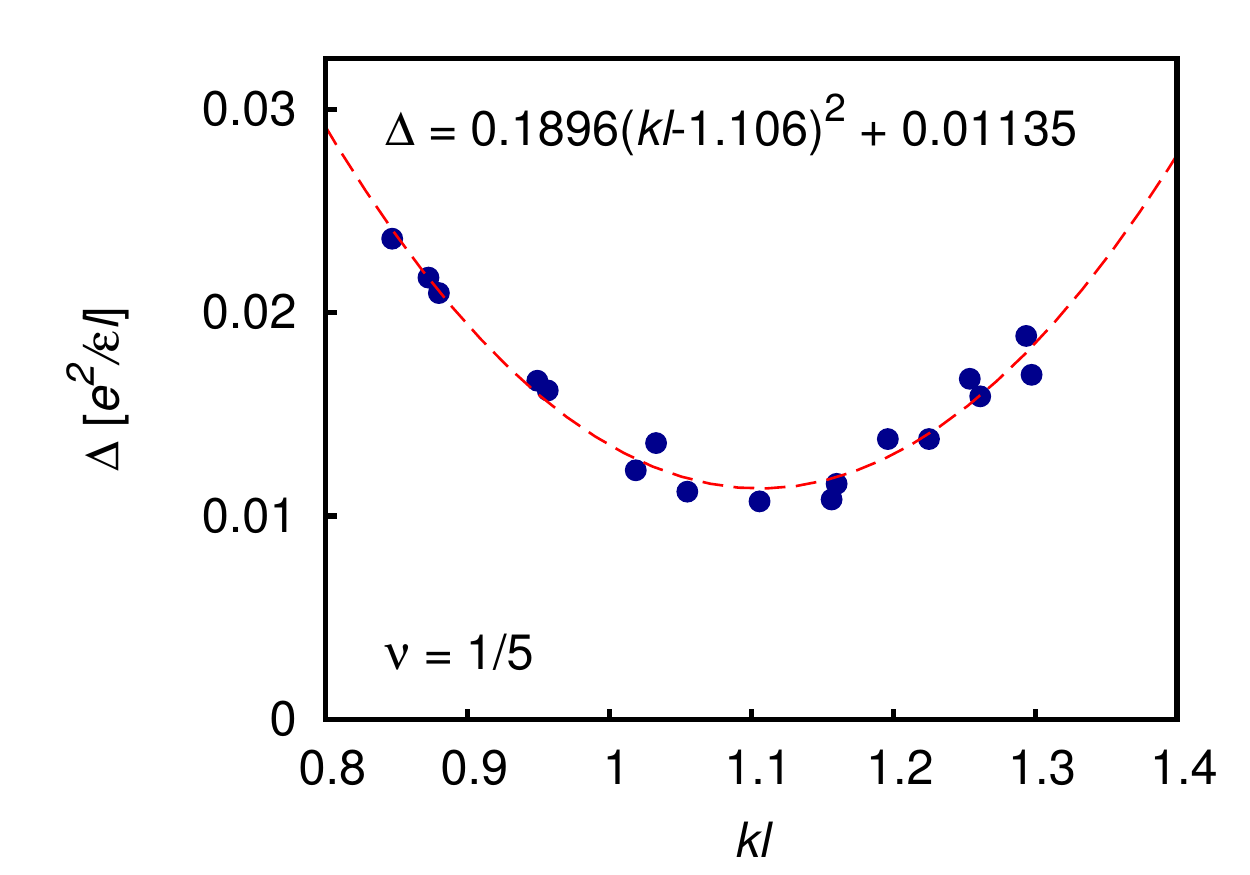}
\includegraphics[width=0.40\textwidth]{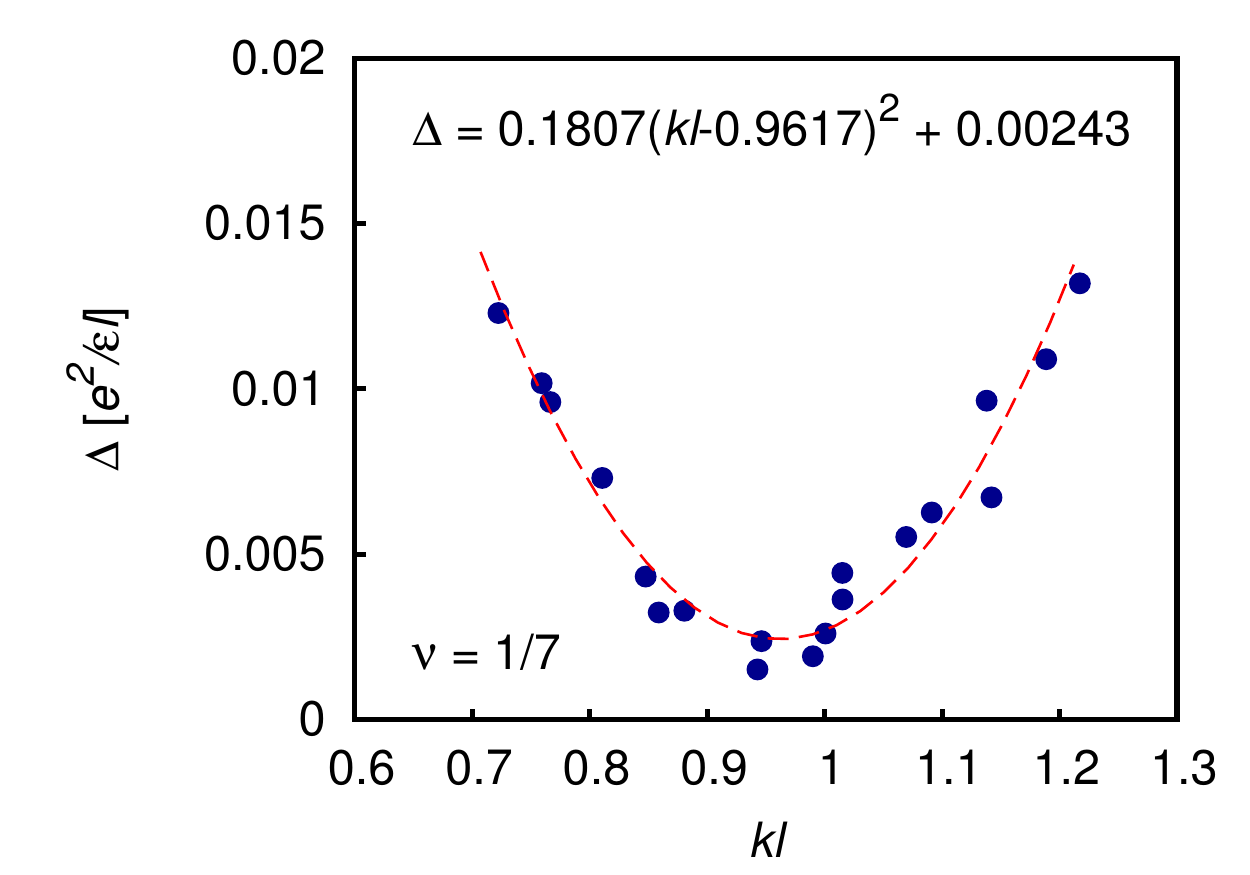}
\end{center}
\caption{Fits to the CF exciton dispersion at $\nu=1/5$ (top panel) and 1/7 (bottom panel) near its minimum energy.  The blue dots are the energies extracted from Ref. [\onlinecite{cfr1}] and the dashed red lines are a least squares fit whose equation is shown in each panel.}
\label{fig:fit}
\end{figure}

\begin{figure}[h]
\begin{center}
\subfloat[$N$=32.]
{
\includegraphics[width=0.37\textwidth]{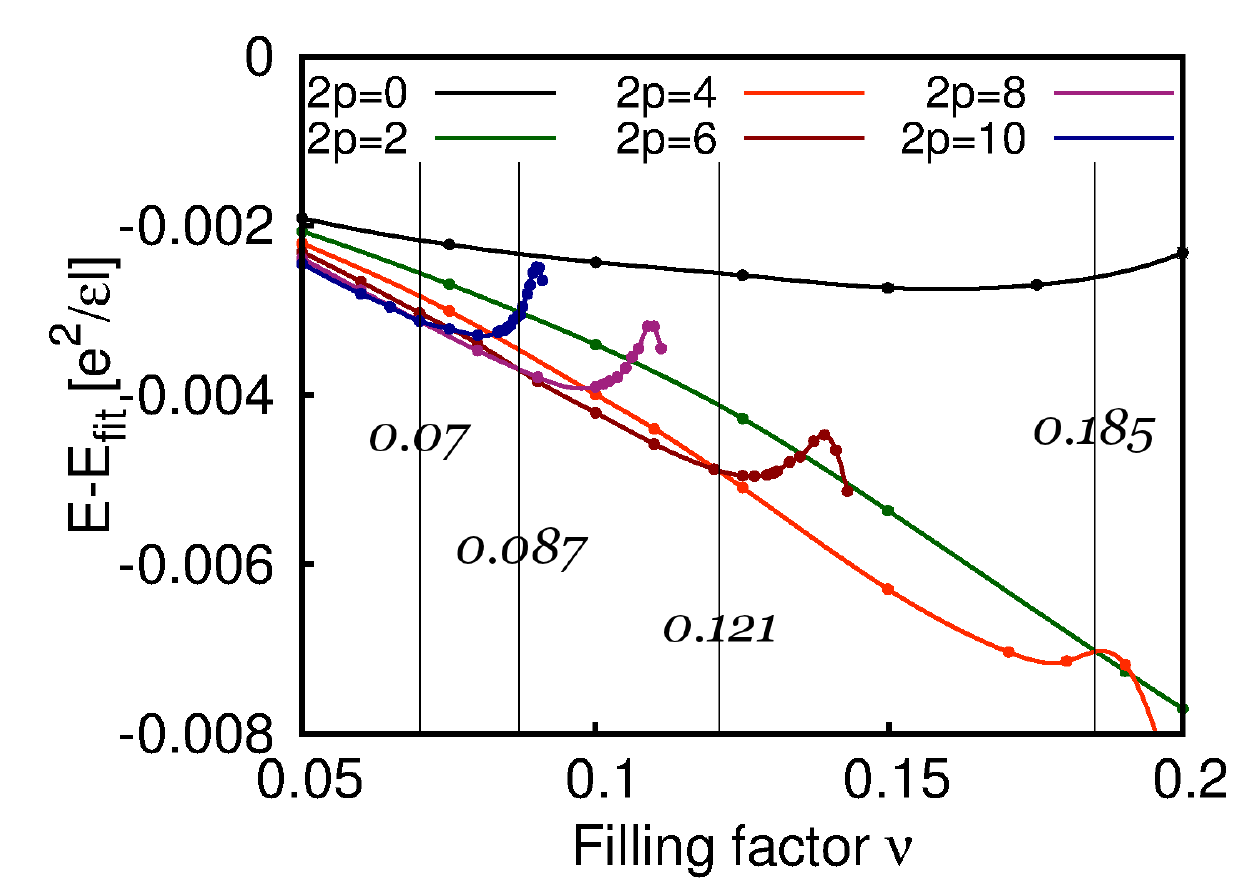}
\label{fig:totE32}
}
\subfloat[$N$=48.]
{
\includegraphics[width=0.37\textwidth]{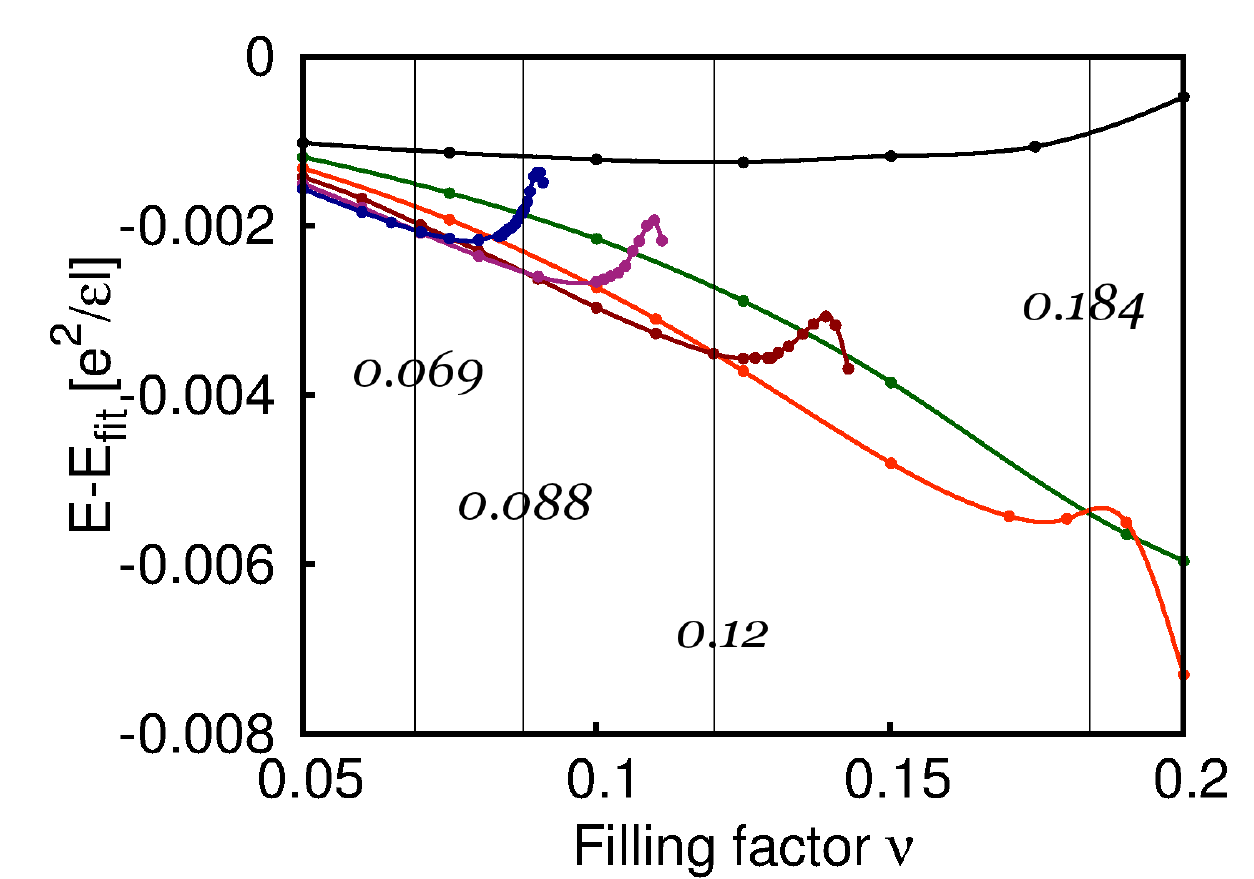}
\label{fig:totE48}
}

\subfloat[$N$=64.]
{
\includegraphics[width=0.37\textwidth]{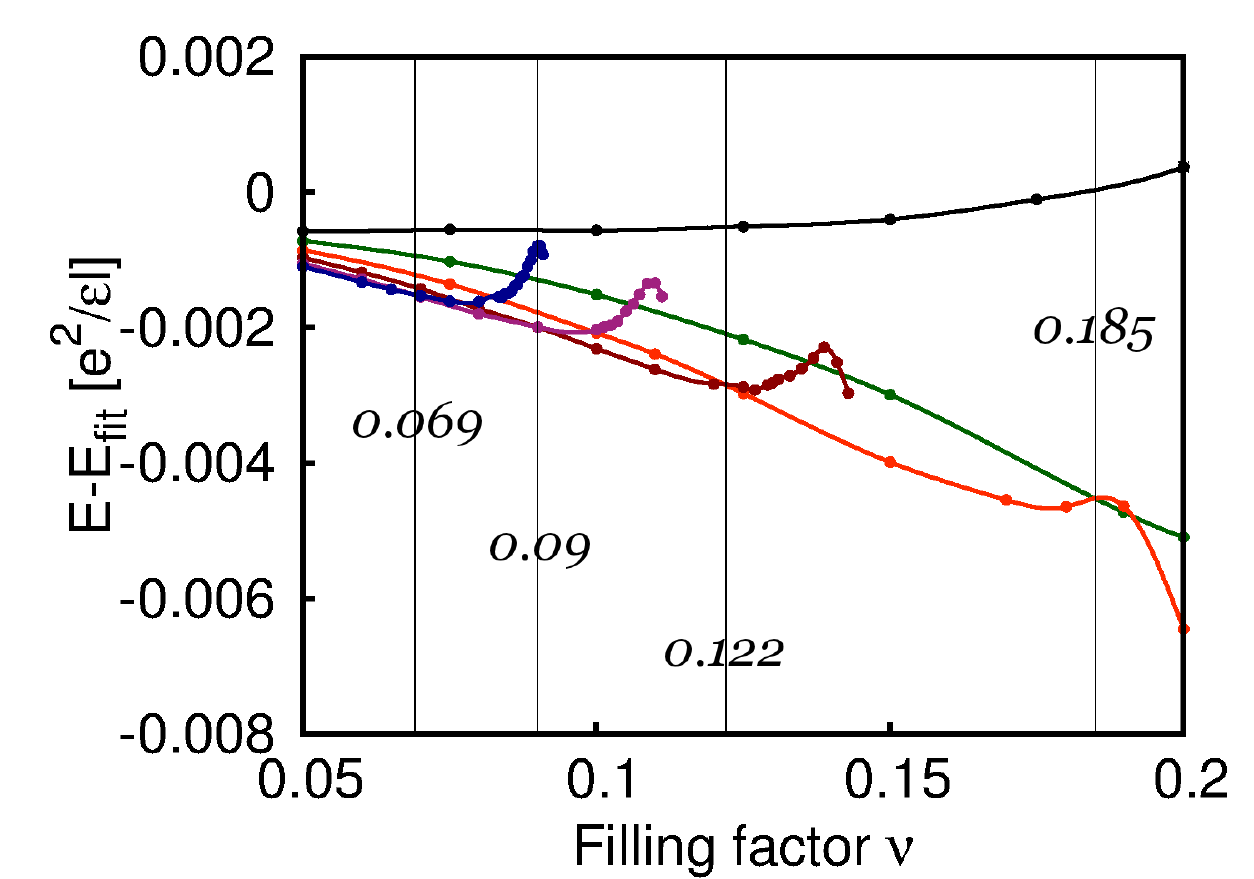}
\label{fig:totE64}
}
\subfloat[$N$=96.]
{
\includegraphics[width=0.37\textwidth]{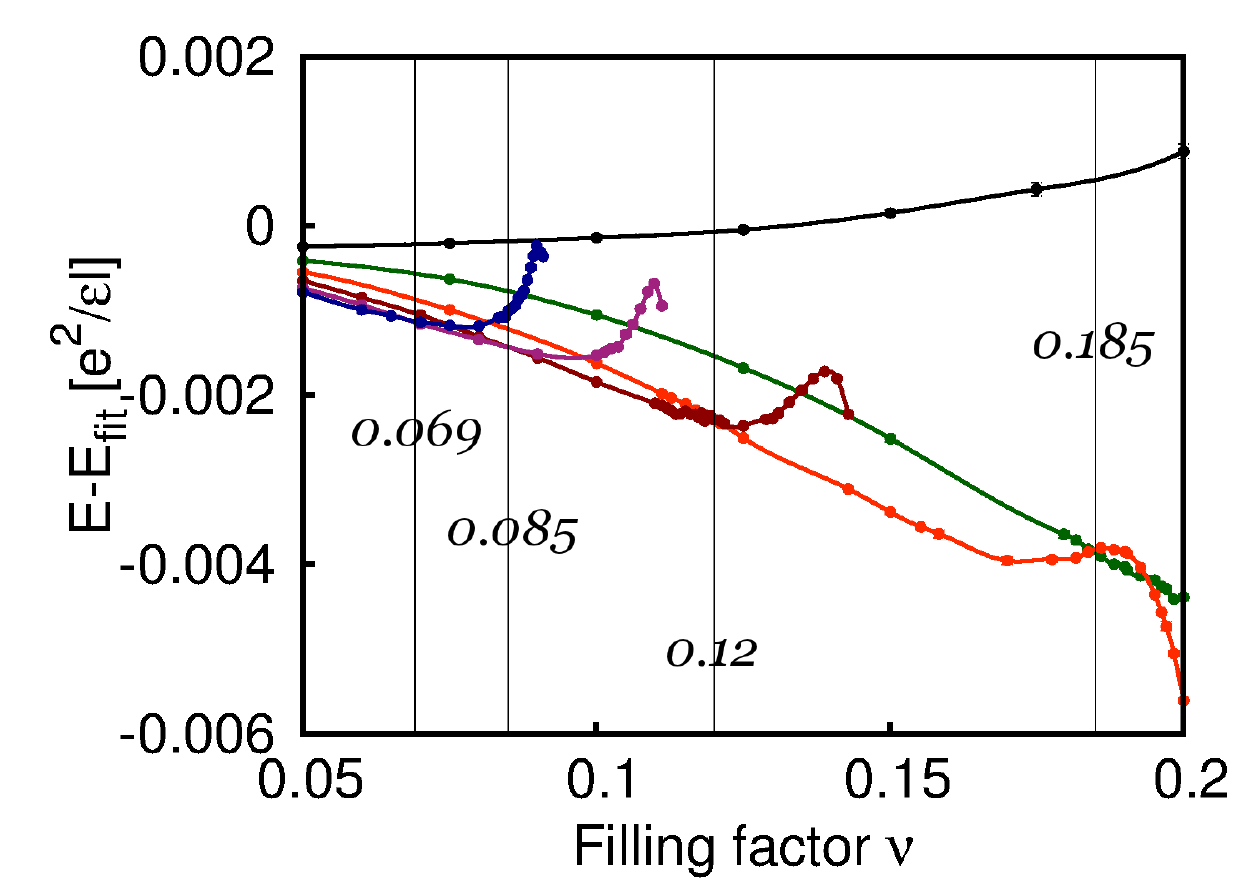}
\label{fig:totE96}
}
\end{center}
\caption{Energy per particle of the type-1 $^{2p}$CFCs for $2p=0,2,4,6,8$ and 10 for several different system sizes. Energies are quoted relative to $E_{\rm fit}=-0.782133\nu^{1/2}+0.2623\nu^{3/2}+0.18\nu^{5/2}-15.1e^{-2.07/\nu}$. The colored dots mark the actual data points and the lines, which are guides for the eyes, are spline interpolations. The thin vertical lines mark the phase boundaries between the type-1 CFCs. We have evaluated a higher density of data points in the filling factor regions with more rapid variations of energy.}
\label{fig:totE}
\end{figure}

\begin{figure}[h]
\begin{center}
\includegraphics[width=0.37\textwidth]{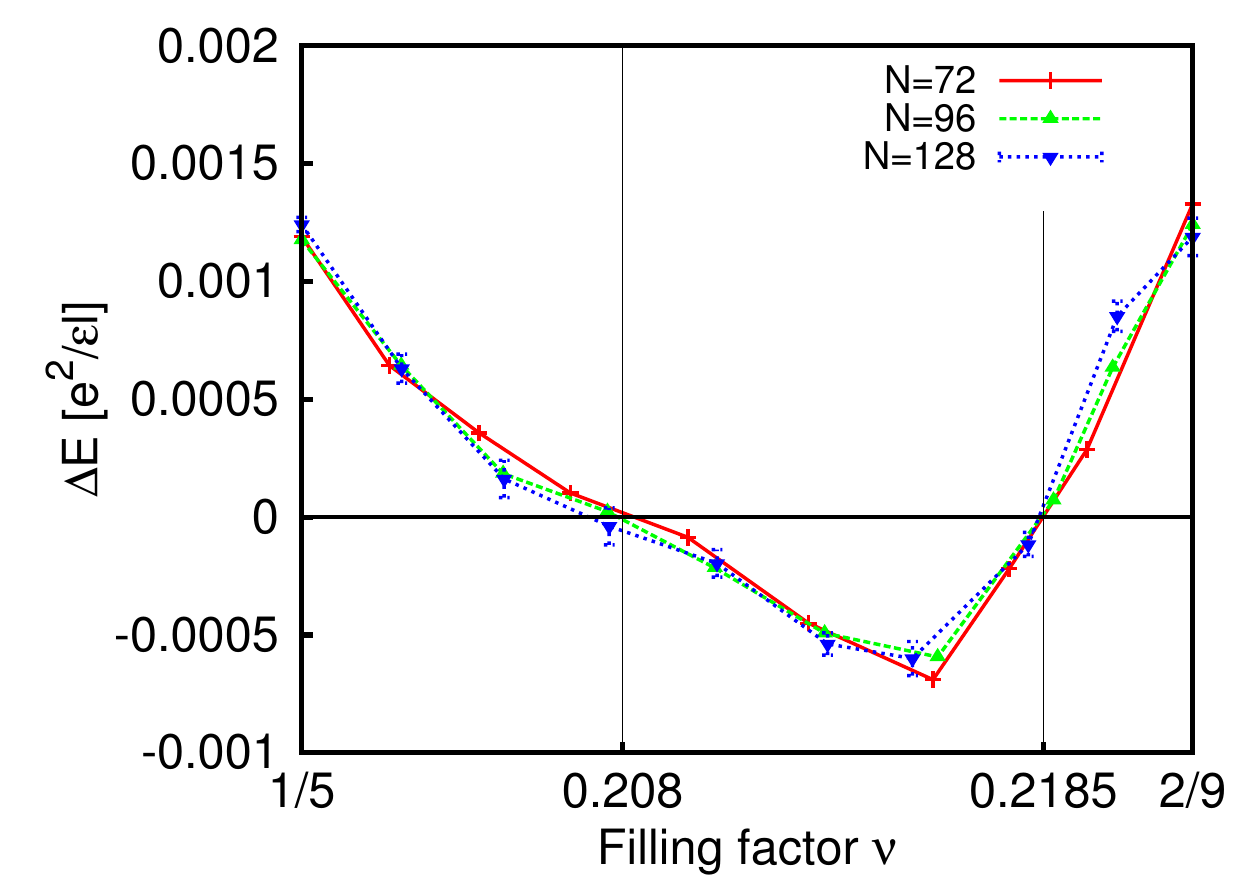}
\end{center}
\caption{Plots of the energy difference of the type-1 $^2$CFC and the FQHE state supporting a type-2 $^4$CFC in the filling factor range $1/5\leq\nu\leq2/9$ for three different system sizes. The uncertainty in the data points is $3\times10^{-5}$.}
\label{fig:ediff}
\end{figure}


\begin{figure}[h]
\begin{center}
\subfloat[$E-E_{\rm MZ}$ for $2p=0$.]
{
\includegraphics[width=0.35\textwidth]{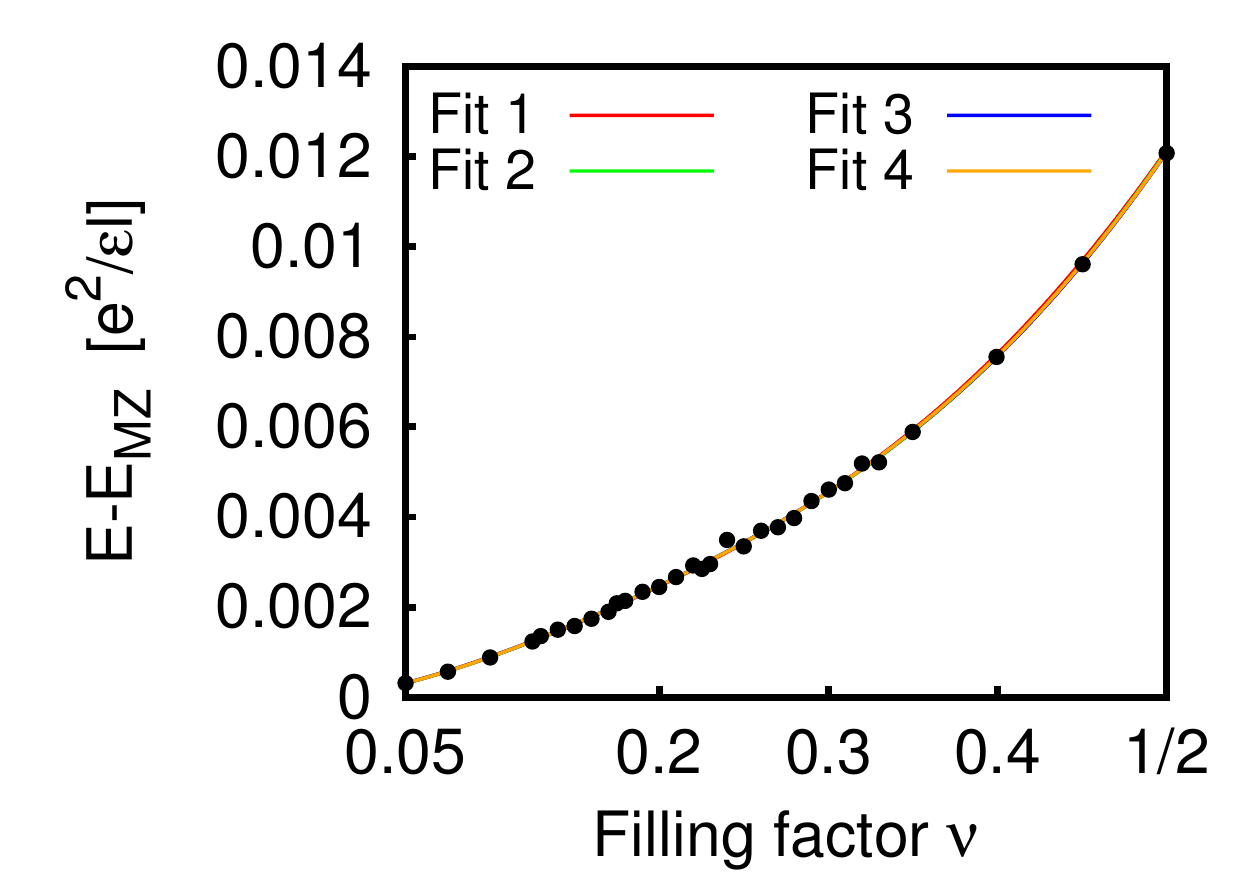}
\label{fig:efitp0}
}
\subfloat[$C^{t}$ for $2p=0$.]
{
\includegraphics[width=0.35\textwidth]{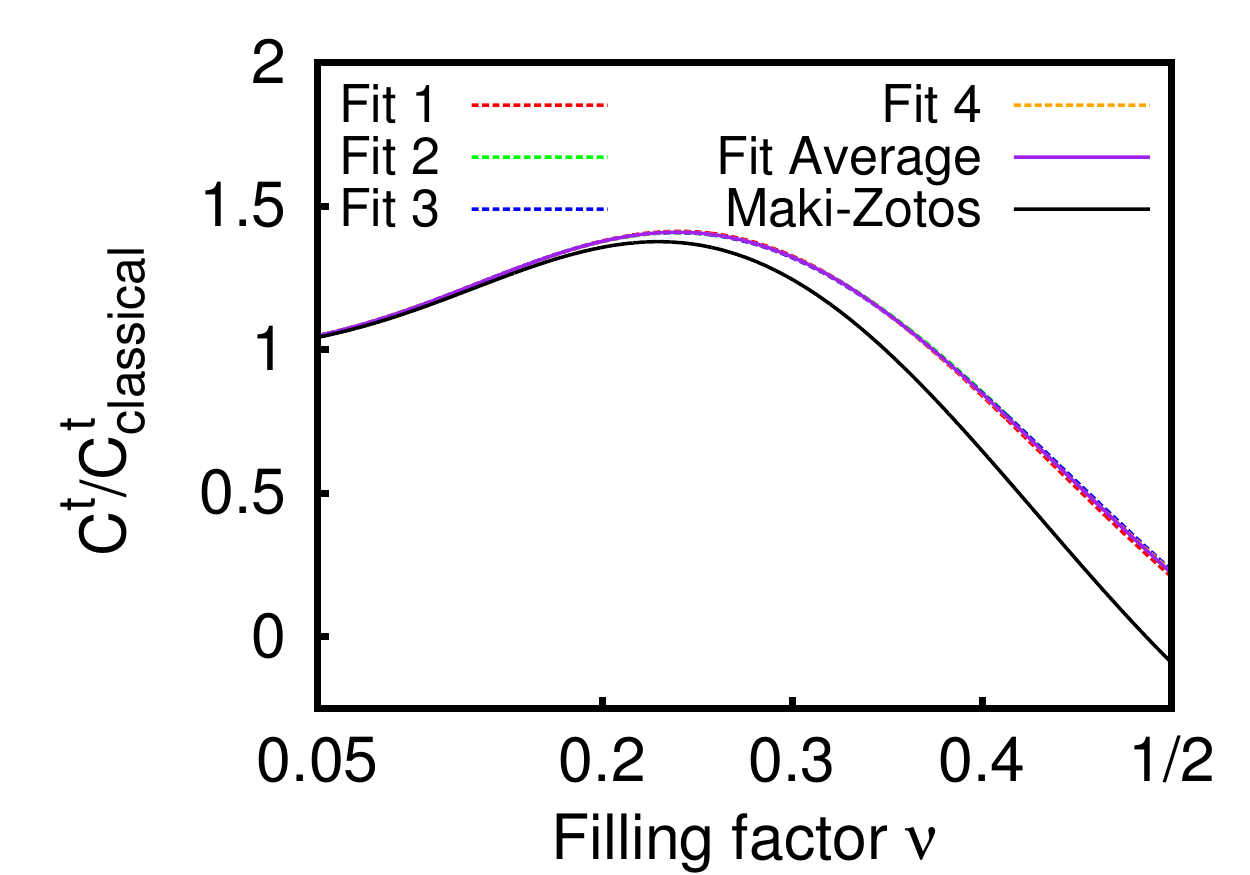}
\label{fig:ctp0}
}

\subfloat[$E-E_{\rm MZ}$ for $2p=2$.]
{
\includegraphics[width=0.35\textwidth]{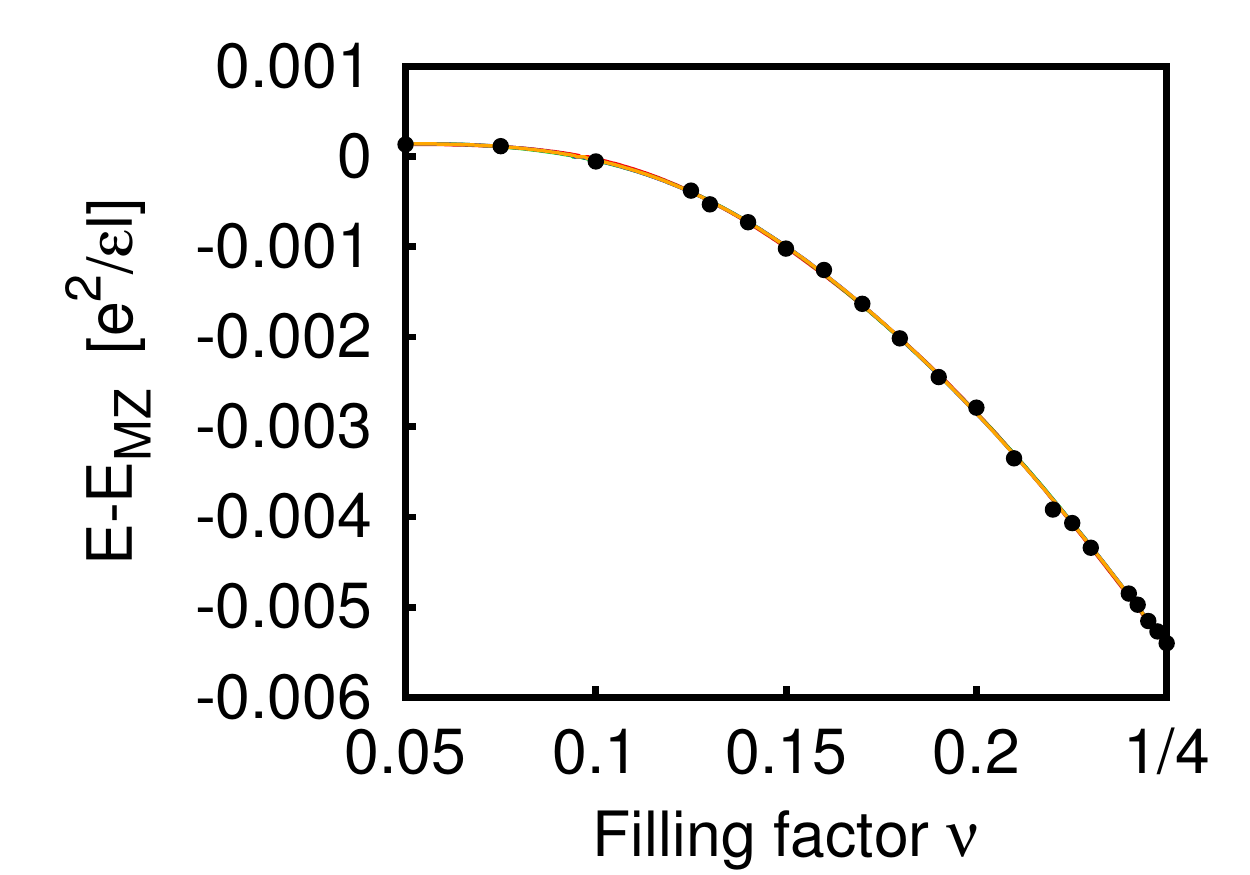}
\label{fig:efitp1}
}
\subfloat[$C^{t}$ for $2p=2$.]
{
\includegraphics[width=0.35\textwidth]{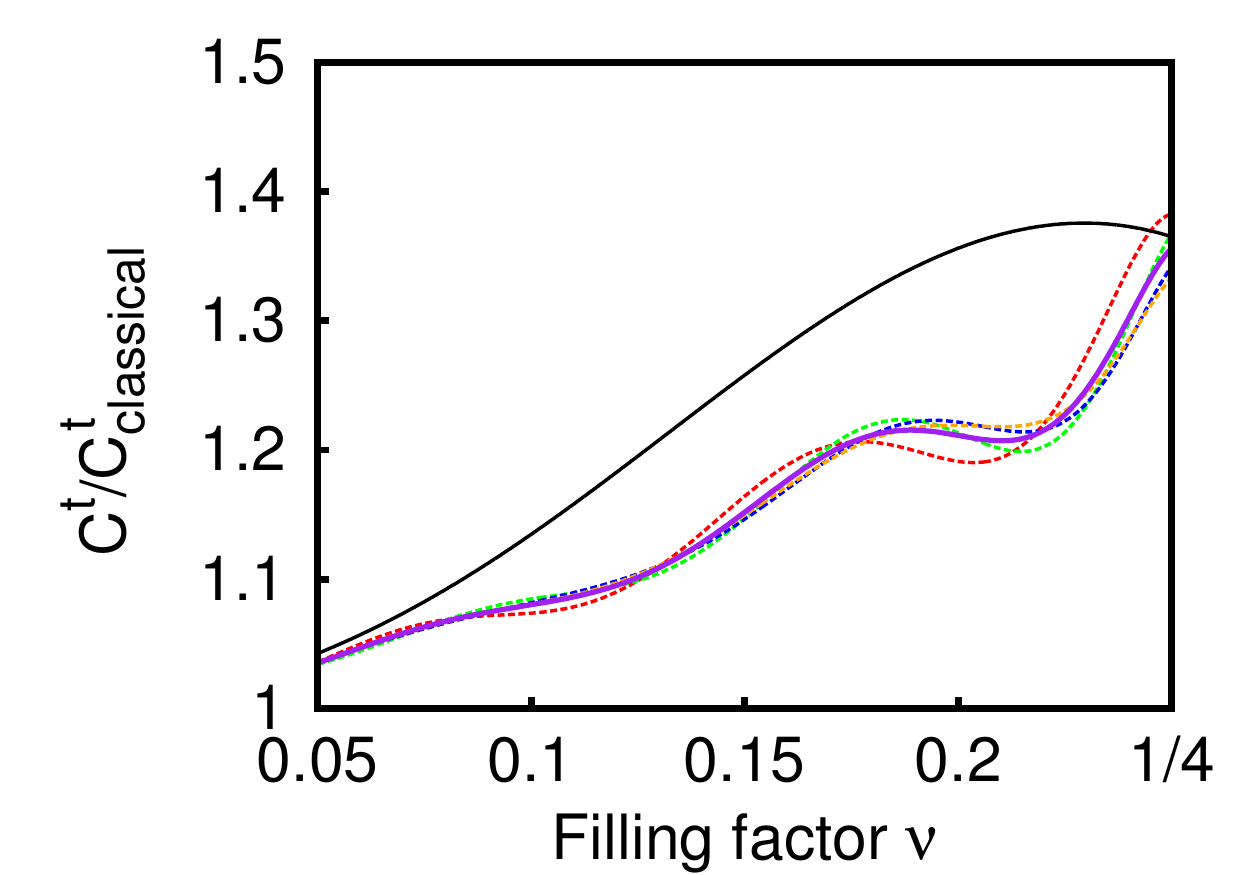}
\label{fig:ctp1}
}

\subfloat[$E-E_{\rm MZ}$ for $2p=4$.]
{
\includegraphics[width=0.35\textwidth]{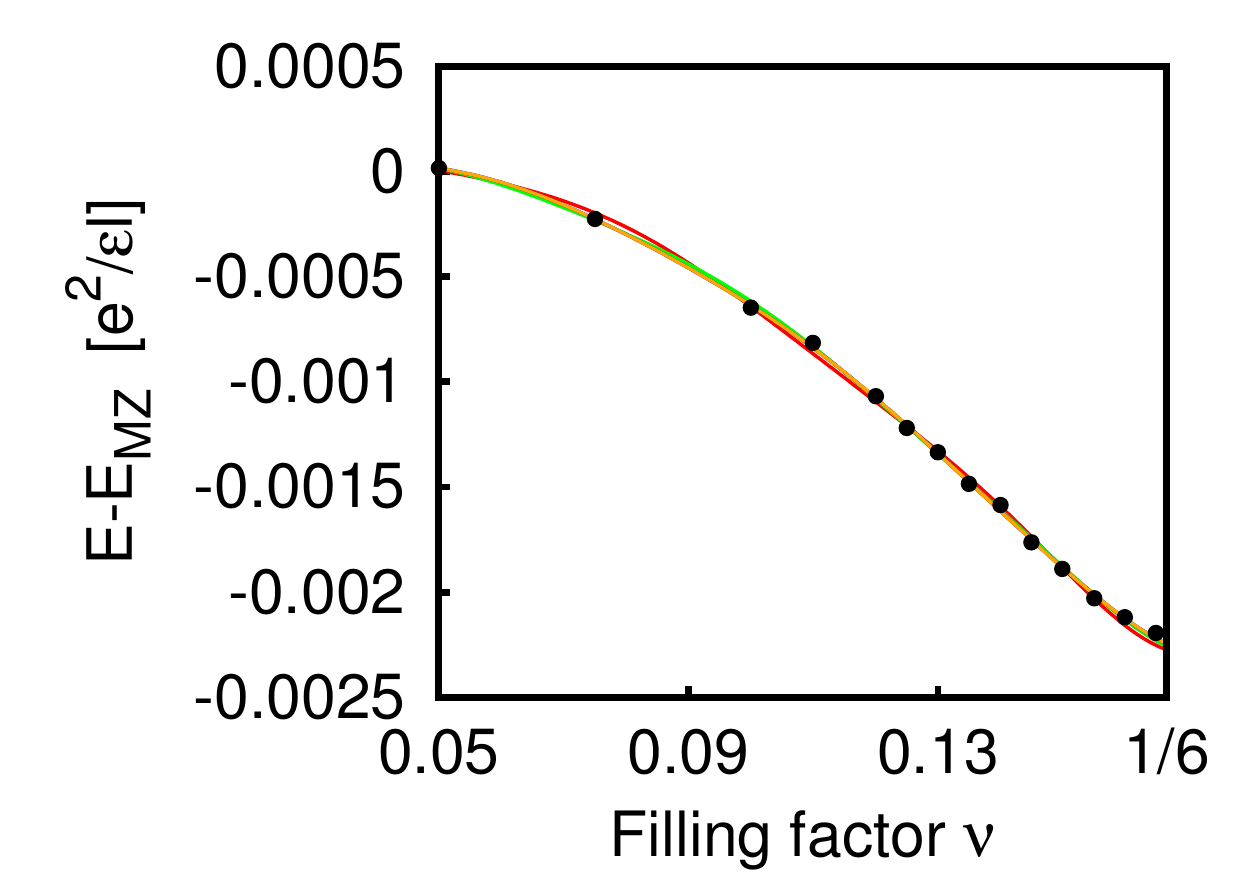}
\label{fig:efitp2}
}
\subfloat[$C^{t}$ for $2p=4$.]
{
\includegraphics[width=0.35\textwidth]{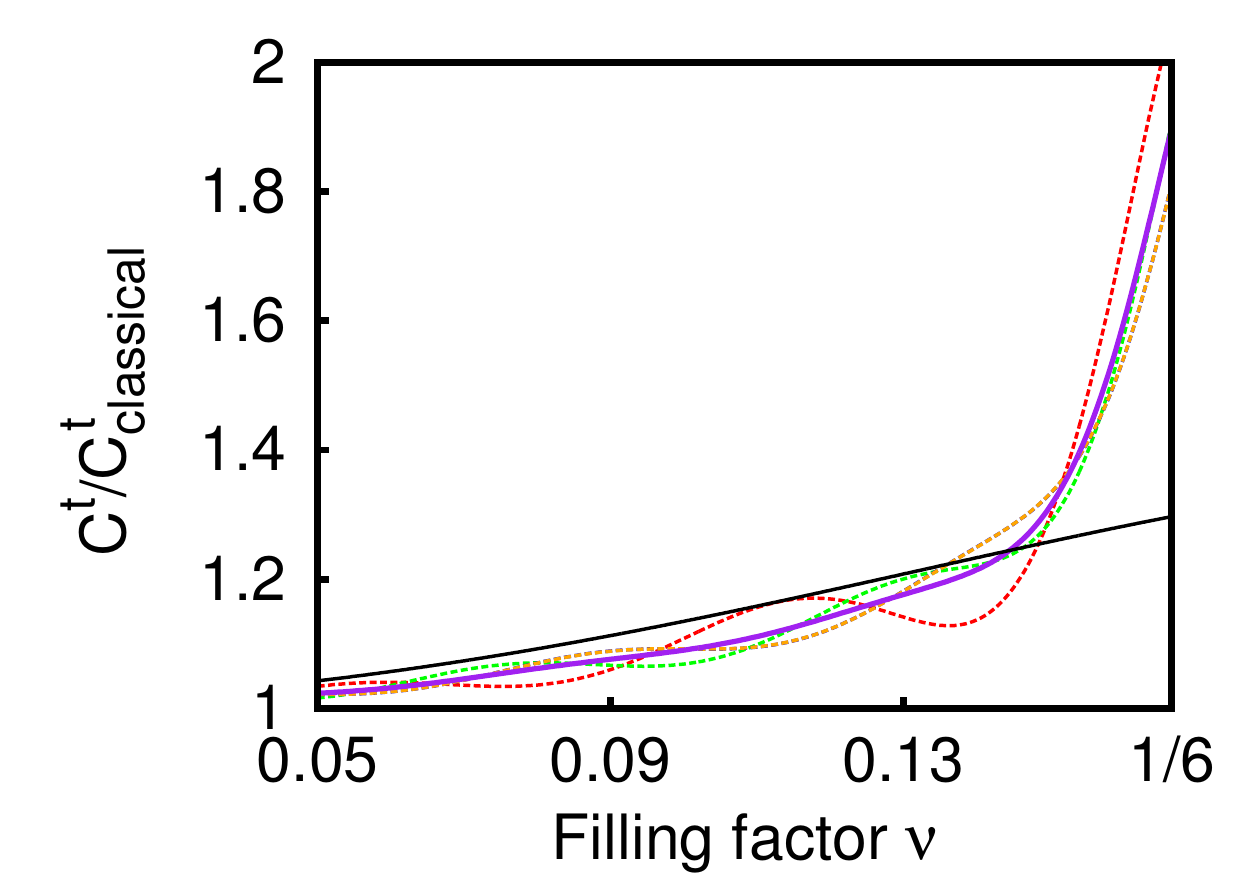}
\label{fig:ctp2}
}

\subfloat[$E-E_{\rm MZ}$ for $2p=6$.]
{
\includegraphics[width=0.35\textwidth]{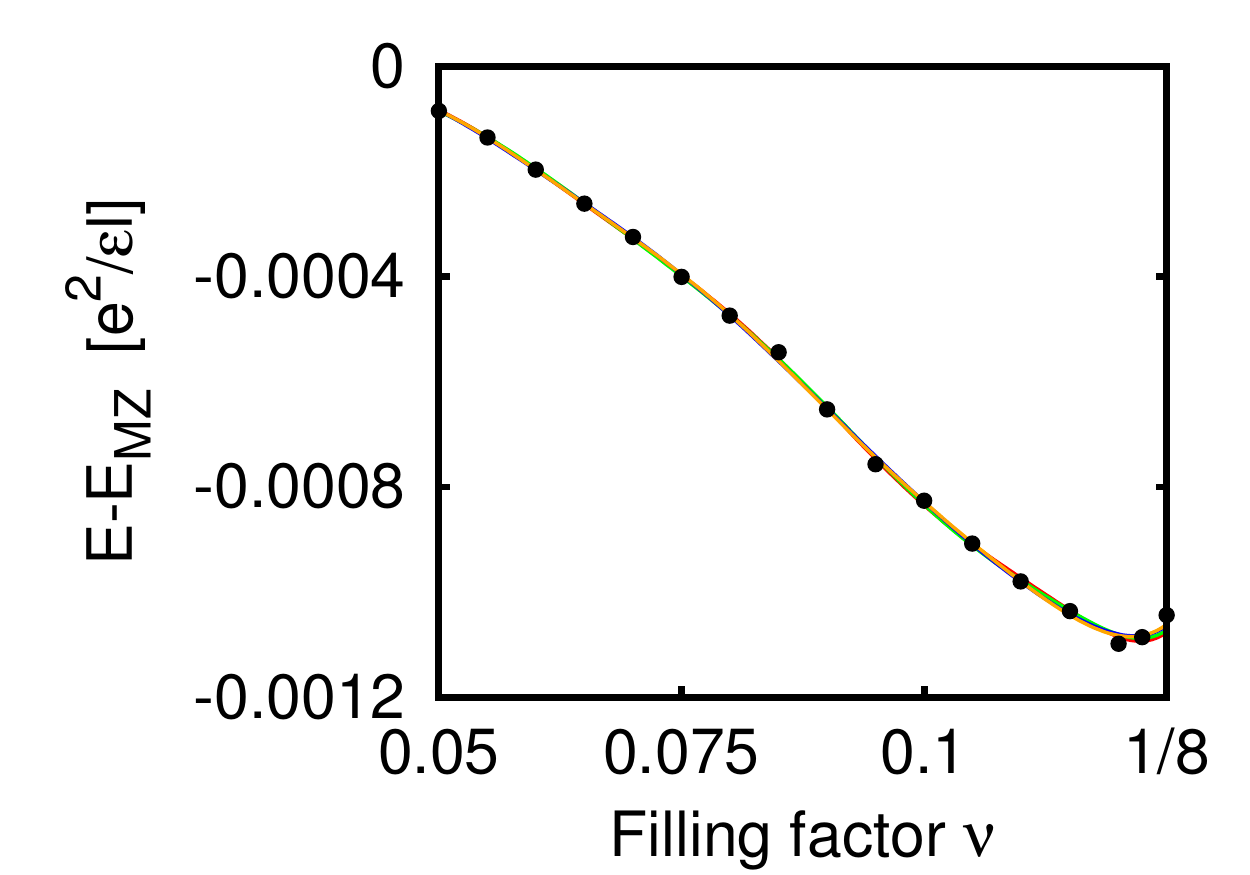}
\label{fig:efitp3}
}
\subfloat[$C^{t}$ for $2p=6$.]
{
\includegraphics[width=0.35\textwidth]{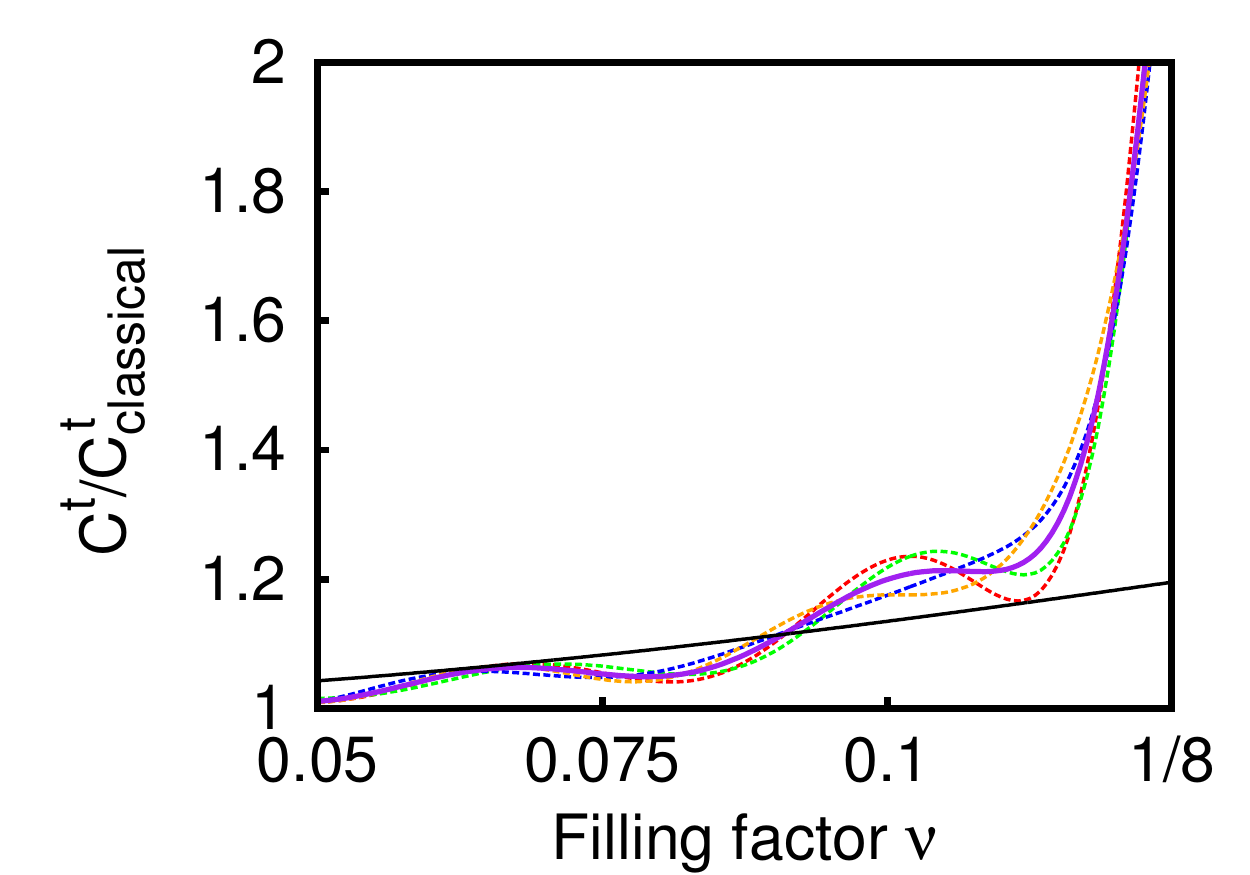}
\label{fig:ctp3}
}
\end{center}
\caption{Panels in the left column show the energies per particle of the $^{2p}$CFCs relative to the Maki-Zotos energy and the lines show the smooth analytical fits made to the data (for details, see text). The panels in the right column show the shear moduli calculated from the smooth analytical fits (dashed curves); the solid purple curve is the average from the four fits shown. The shear modulus calculated using the two-body Maki-Zotos interaction, shown by solid black line, is included for reference. All energies are calculated for a system with $N=128$ particles.}
\label{fig:ct}
\end{figure}

\clearpage

\begin{figure}[h]
\includegraphics[width=0.55\textwidth]{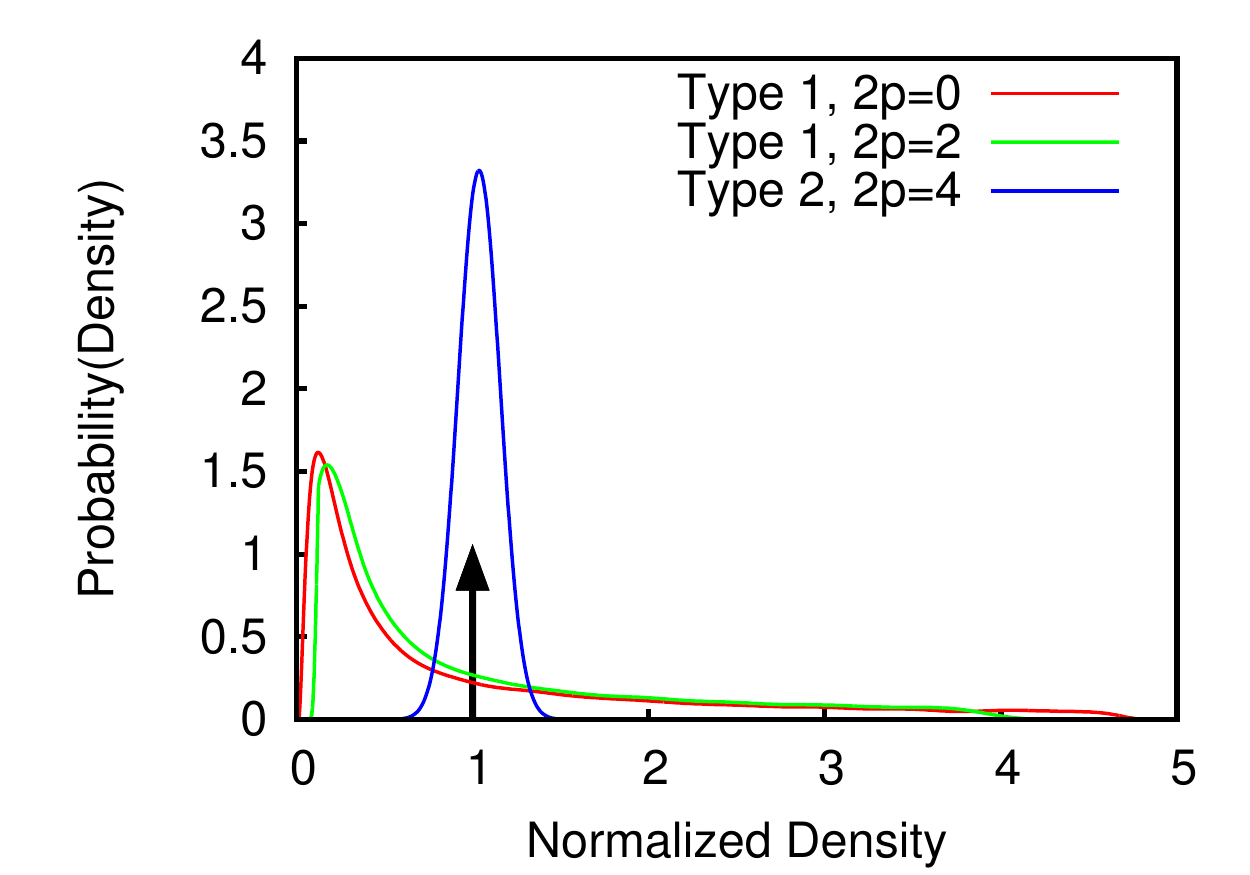}
\caption{Plot of the normalized probability of the density vs. density within a Wigner-Seitz cell on the sphere for type-1 and type-2 CFCs at $\nu=0.215$. The black arrow is a delta function centered at a density of 1 and represents the probability of the uniform-density FQHE liquid.}
\label{fig:pdens}
\end{figure}

\end{appendix}

\clearpage
\end{widetext}

\end{document}